\documentclass[12pt,preprint]{aastex}
\usepackage{lscape}

\newcommand{\myemail}{mizuta@MPA-Garching.MPG.DE}

\slugcomment{accepted for publication in ApJ}

\shorttitle{Collimated Jets and Expanding Outflows}
\shortauthors{Mizuta et al.}

\begin{document}

\title{Collimated Jet or Expanding Outflow:
Possible Origins of GRBs and X-Ray Flashes}

\author{Akira Mizuta\altaffilmark{1,2,3},
Tatsuya Yamasaki\altaffilmark{1},
Shigehiro Nagataki\altaffilmark{1,4}, and
Shin Mineshige\altaffilmark{1}\\
}

\altaffiltext{1}{Yukawa Institute for Theoretical Physics, Kyoto
University, Oiwake-cho Kitashirakawa, Sakyo-ku, Kyoto, 606-8502 Japan.}
\altaffiltext{2}{Max-Planck-Institute f\"ur Astrophysik
Karl-Schwarzschild-Str. 1, 85741 Garching, Germany}
\altaffiltext{3}{E-mail:\myemail}
\altaffiltext{4}{KIPAC, Stanford University, P.O.Box 20450, MS 29,
Stanford, CA, 94309, USA}

\begin{abstract}
We investigate the dynamics of an injected outflow 
propagating in a progenitor
in the context of the collapsar model 
for gamma-ray bursts (GRBs) through two dimensional 
axisymmetric relativistic hydrodynamic simulations.
Initially, we locally inject an outflow near the center
of a progenitor.
We calculate 25 models, in total, 
by fixing its total input energy to be $10^{51}\mbox{ergs\ s}^{-1}$
and radius of the injected outflow to be $7\times 10^7$ cm 
while varying its bulk Lorentz factor, $\Gamma_{0} = 1.05\sim 5$,
and its specific internal energy, $\epsilon_0/c^2 = 0.1\sim 30$ 
(with $c$ being speed of light).
The injected outflow propagates in the progenitor
and drives a large-scale outflow or jet.
We find a smooth but dramatic transition from a
collimated jet to an expanding outflow among calculated models.
The opening angle of the outflow ($\theta_{\rm sim}$) is sensitive 
to $\Gamma_0$; we find $\theta_{\rm sim} < 2^\circ$ for $\Gamma_0 \gtrsim 3$.
The maximum Lorentz factor is, on the other hand,
sensitive to both of $\Gamma_0$ and $\epsilon_0$;
roughly $\Gamma_{\rm max} \sim \Gamma_0 (1+\epsilon_0/c^2)$.
In particular, a very high Lorentz factor
of $\Gamma_{\rm max} \gtrsim 100$ is achieved in one model.
A variety of opening angles can arise 
by changing $\epsilon_0$,
even when the maximum Lorentz factor is fixed.
The jet structure totally depends on $\Gamma_0$.
When $\Gamma_0$ is high,
a strong bow shock appears and generates a back flow.
High pressure progenitor gas heated by the bow shock 
collimates the outflow to form a narrow, relativistic jet.
A number of internal oblique shocks within the jet
are generated by the presence of the back flow and/or shear
instability.
When $\Gamma_0$ is low, on the contrary,
the outflow expands soon after the injection,
since the bow shock is weak and thus
the pressure of the progenitor gas 
is not high enough to confine the flow.
Our finding will explain a smooth transition between the GRBs,
X-ray rich GRBs (XRRs) and X-ray Flashes (XRFs) by the same 
model but with different $\epsilon_0$ values.
\end{abstract}


\section{INTRODUCTION}
The GRBs are to our best knowledge the most energetic phenomena 
in the Universe.  So far intense efforts have been made both on
the observational and theoretical grounds toward 
understanding of their natures, but
their origins still remains to be investigated.
One of the most important finding recently is
that GRBs are involved with relativistic collimated flows.
To account for the observations, an extremely high 
bulk Lorentz factor, typically more than 100,
is required \citep{Rees92}.

GRBs are known to be composed of two classes:
long-duration GRBs (with duration being longer than a few seconds) 
and short-duration GRBs (with duration being less than a second).
At least, some of the long-duration GRBs are known to be
associated with supernovae (SNe).  
Good examples of GRB-SN connection are
GRB980425/SN1998bw \citep{Galama98} and GRB030329/SN2003dh
\citep{Stanek03,Uemura03,Hjorth03,Price03}.
These provide strong evidences that the central engines of
(at least part of) the long-duration GRBs are SNe.
Such association was theoretically predicted
by \citet{Woosley93} and \citet{Paczynski98}.
A signature of a supernova contribution is actually
found in the afterglow spectra of these GRBs.
These supernovae are categorized in the type Ic
whose progenitor has lost its hydrogen and helium envelope
when the core-collapse occurs.
Another possible GRB associated with SN is GRB021211.
A strong absorption feature is observed
in the spectrum of the afterglow \citep{DellaValle03}.
They concluded that the absorption is due to CaII
which is synthesized by the associated supernova explosion.
Further, the discovery of
the host galaxy of the long duration GRBs being
star forming galaxies \citep{Bloom02,LeFloch03,Tanvir04}
also strengthens the idea of strong GRB-SN connection.

It has been suggested that even supernovae,
in which no associated GRB was found, might have
a link with GRB.
For example, the peculiar SN2002ap recorded
a high velocity component of $0.23 c$ and huge
kinetic energy of the jet of $5\times 10^{50}$ ergs, at least.
These values are similar to those of GRBs,
indicating a similar explosion mechanism of SN2002ap to 
that of GRBs.
\citet{Totani03} concluded that SN2002ap is one example
of the supernovae which failed to make a GRB.  

Recently, a number of X-ray-rich GRBs (XRRs) and XRFs,
very similar phenomena to GRBs but with significantly
lower peak energy, have been successively discovered 
thanks to the good performance of HETE-2 
(see \citet{Heise01} and \citet{Sakamoto05}).
Interestingly, the event rates of XRRs and XRFs are 
similar to that of the long duration GRBs \citep{Sakamoto05}.
The origin of these events is poorly understood, but
similar burst properties of GRB, XRRs, and XRFs 
except for peak energy
leads to an idea that they all might have the same
origins but with different viewing angles (Nakamura 2000)
or with variable opening angles (Lamb et al. 2005).

In this paper, we elucidate the theory of XRFs, XRRs, 
and GRBs in the context of the so-called collapsar model.
The collapsar is a death of a massive star 
in the last stage of the stellar evolution.
The collapsar model for GRBs was proposed by Woosley 
(1993; see also MacFadyen \& Woosley 1999),
for a central engine of GRBs.
In this model, strong outflows, or jets,
emerge from deeply inside the collapsar and propagate
into the interstellar medium (ISM), producing gamma-ray bursts.
\citet{MacFadyen99} performed two dimensional hydrodynamic simulations
based on this model.
Assuming annihilation of neutrino and anti-neutrino,
they deposited thermal energy around the center of the core
which had been collapsed and become a black hole.
Their initial mass density profile is very flattened
due to rotation of the progenitor.
The gas around the center, where the high thermal energy is deposited,
expands and forms very collimated outflow; i.e., a ``jet''.
The outflow successfully became a bipolar outflow.
Unfortunately, however, their calculations were not relativistic one,
so the relativistic effects which are important to understand
GRBs were not included.

Since the mass density of the progenitor is quite high,
it is not a trivial issue
whether or not the formed outflow can always keep collimated
structure and break out from the progenitor surface as a jet.
It has been pointed out through the AGN jet simulations 
that the multi-dimensional effects are so important
for the dynamics of the ``light jet'' into some dense gas
(see, e.g., \citet{Mizuta04}, and references therein).
Here, light jet stands for the jet,
the mass density of which is smaller than that of the ambient gas.
At least two dimensional hydrodynamic calculations
are indispensable to investigate the outflow propagation 
and its dynamics inside and outside the progenitor.

Multi-dimensional, relativistic hydrodynamic simulations 
have been so far performed by several groups
in the context of collapsar model \citep{Aloy00,Zhang03,Zhang04,Umeda05}.
\citet{Aloy00}, for example, performed relativistic hydrodynamic 
simulations based on the model by \citet{MacFadyen99}.
They have found that
a bipolar flow is created and it breaks out from the progenitor.
Interestingly, the maximum Lorentz factor of about 40 has been 
achieved, when the jet breaks out of the progenitor.
Another type of relativistic hydrodynamic simulations
of the outflow have also been 
performed by \citet{Zhang03}, \citet{Zhang04}, and \citet{Umeda05}.
They modeled a mainly very hot jet, i.e.
an initially thermal energy dominated jet.
Injected jets from the computational boundary,
which is assumed to be very close to the collapsed center of the progenitor,
always successfully propagate in the progenitor keeping
good collimation and break out the progenitor.

It is still open question, however, whether
a collimated outflow emerging from the center of the progenitor
can always break out or not.
Although several provenance studies have demonstrated
successful propagation throughout the progenitor and
breakout of the input outflow, they might have assumed
unrealistically large energy input in the initial condition. 
We should be aware that
the formation mechanism of the outflow from the center 
of the collapsed progenitor has been poorly understood.
In other words,
we still do not know the physical conditions
(density, thermal energy, kinetic energy, magnetic energy, 
opening angle etc) for generating outflows.
Further, it is not completely clear yet
what discriminates between SNe associated with GRBs
and those without GRB association.
The connection between XRFs and GRBs is another important issue.
A key factor may be attributed to 
the different dynamics of the outflow propagation.

Motivated by these questions
we perform series of relativistic hydrodynamic simulations
of the outflow propagation in the progenitor and ISM.
In these simulations, we fix
the total input energy power of $10^{51} \mbox{ergs s}^{-1}$
but vary the bulk Lorentz factor ($\Gamma_0$)
and the specific internal energy of the initial outflow
($\epsilon_0$ which excludes the rest mass energy).
We discuss what types of outflow can emerge from the central system
for a wide range of parameters, $\Gamma_0$ and $\epsilon_0$.

This paper is organized as follows.
In Sec. \ref{method} we introduce our model
and then explain the numerical methods and the
initial background and outflow conditions.
The results of the numerical simulations are presented
in Sec. \ref{results}, where
we will demonstrate the emergence of
two distinct types of outflows:
a collimated jet and an expanding outflow.
We then discuss the dynamics and structure of the outflow,
focusing on the distinctions between the two types of flows
in Sec. \ref{discussion}.
The final section is devoted to conclusions.

\section{NUMERICAL METHOD AND CONDITIONS}
\label{method}
\subsection{Our Model: Progenitor and Injected Outflow}

Our model is based on the collapsar model
(see, e.g., Woosley 1993).
According to this model
the release of the gravitational energy is
the main energy source of the energetic phenomena.
This model explains the formation of an outflow as follows:
When an iron core of the massive star collapses,
a system consisting of a black hole or a proto-neutron star
and an accretion disk is formed at the center of the progenitor.
Some fraction of the collapsing gas can produce an outflow,
although the formation mechanism of this outflow is not 
fully understood yet, as in the case of AGN jets.
Annihilation of neutrinos and anti-neutrinos
emitted from the accretion disk or via MHD process 
is a possible scenario to generate such outflows.
The progenitor is expected to spin rapidly.
The gas along the rotational axis can freely fall into the center
while the gas along the equatorial plane only gradually falls
because of stronger centrifugal force.
As a result, a tenuous regions are created along the rotational axis.
After the core collapse, the outer envelopes begin to fall freely. 
The free fall timescale of the envelopes is longer
than the dynamical time scale for the jet to propagate within the 
progenitor and to hit its surface. 
After the formation of the outflow,
it should propagate in the progenitor and
break out the surface of the progenitor into ISM.
Finally the outflow is observed as a GRB and an afterglow.

We adopt the radial mass profile from \cite{Hashimoto95},
assuming that the progenitor is spherically symmetric
when the iron core was collapsed
(see however \citet{Petrovic05,Yoon05,Woosley06} for recent calculations
on the effects of the angular momentum
and magnetic fields).
At least near the core, the profile  along the equatorial axis
is very similar to that
used by \cite{MacFadyen99} and \citet{Aloy00}.
The progenitor had a mass of about 40 solar masses in the main sequence
and has 16 solar masses in the pre-supernovae stage.
The hydrogen envelope has already been lost
by the stellar wind.
Although our progenitor includes helium envelope
which would produce type Ib SN,
main properties of the outflow dynamics, such as,
collimated jet or expanding outflow described below,
strongly depend on the mass profile around the injection point,
i.e. silicon, carbon, and oxygen envelopes.
Our main conclusion is not affected whether we adopt the progenitor
which includes helium envelope or not.

\begin{figure}
\epsscale{.80}
\plotone{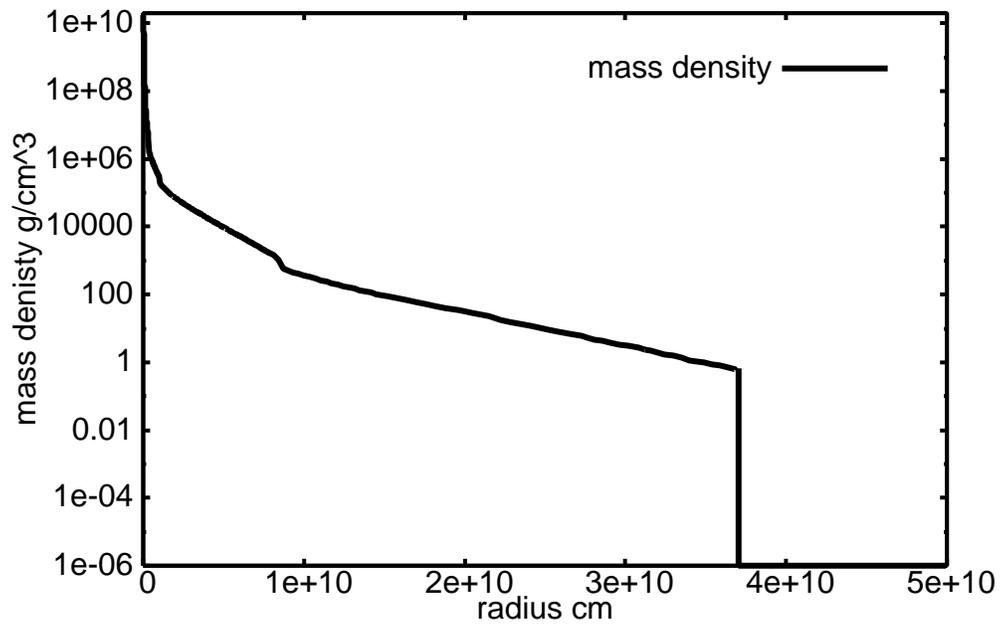}
\caption{The adopted radial mass density profile of the progenitor
 from the center (left) of the core to the surface (right).
We adopt the profile at $r>2\times 10^8\mbox{cm}$ as our initial
mass profile, assuming spherical symmetry.
This is equivalent to assuming that
a mass of about two solar masses has collapsed.}
\label{1Ddensity}
\end{figure}

\begin{figure}
\epsscale{.80}
\plotone{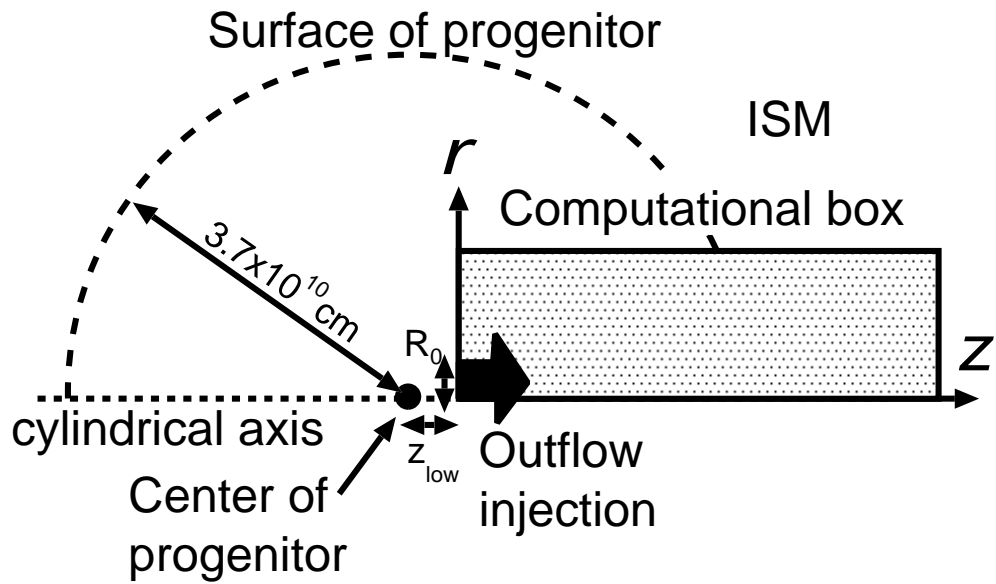}
\caption{A schematic figure of the progenitor and
computational box throughout the present study.
The radius of the injected outflow ($R_0$) is $7\times 10^7$ cm.
The distance between the center of the progenitor and
the lower-left corner of the computational box ($z_{\rm low}$) is
$2\times 10^8$ cm or $2\times 10^7$ cm.}
\label{condition_fig}
\end{figure}

Figure \ref{1Ddensity} shows the radial mass density profile
from the center to the surface of the progenitor.
The mass density decreases from $10^{10}\ \mbox{g cm}^{-3}$ to
$1\ \mbox{g cm}^{-3}$
from the center to the surface located at 
3.7 $\times 10^{10}$cm from the center.
The lower boundary of the computational domain ($z_{\rm low}$) is 
set to be $2\times 10^8$ cm
from the center of the progenitor.
The center of the progenitor is set to be $z=0$
(see Figure \ref{condition_fig} for a schematic view 
 of the progenitor and the location of the computational box).
Note that the total mass of the progenitor
within this radius is about two solar masses; that is,
here we postulate the situation that material
of about two solar masses collapses towards the center
and forms the system of a proto-neutron star or black hole
and a surrounding accretion disk.
This distance between the center of the progenitor and
the computational lower boundary is the same as that adopted
by \citet{Zhang03}.
We will also calculate an additional case study,
in which the lower boundary is set to be 
at $2\times 10^7$ cm from the center, following \citet{Aloy00},
in which thermal energy is deposited as a driver of the outflow.

Since the pressure becomes very high due to the strong bow shock,
the pressure of the progenitor can be negligible.
The initial thermal energy in the progenitor is set to be very low.

The formation mechanism of the outflows around the core
is not fully understood yet.
We assume injection of an initial outflow from the lower boundary
and that the direction of the initial outflow
is parallel to that of the cylindrical ($z$) axis.
This method is basically the same as that adopted by \cite{Zhang03} 
\citet{Zhang04} and \citet{Umeda05},
although the injected outflows calculated by \citet{Zhang03} and
\citet{Umeda05}
had a finite opening angles and were not initially parallel flow.
In the present study
the radius ($R_0$) and power ($\dot{E}_0$)
of the injected outflow is fixed to be
$R_0=7\times 10^{7}\mbox{cm}$ and
$\dot{E}_0=10^{51}\mbox{ergs s}^{-1}$,
respectively.
In this paper the subscript 0 
stands for the values of the injected outflow
and $\dot{E}_0$ does not include the rest mass energy.
The energy flux ($\dot{E}_0/\pi {R_0}^2$)
does not depend on radius.
The net injected energy during the first ten seconds 
amounts to $10^{52}$ ergs.  This values is
close to the explosive energy of SN1998bw and SN2003dh
and is higher than the normal explosive energy of
$\sim 10^{51}$ ergs \citep{Iwamoto98,Woosley99,Hjorth03}.
Such energetic supernovae are sometimes categorized as hypernovae.

Two more parameters are necessary to specify the outflow condition.
The bulk Lorentz factor ($\Gamma_0$) and specific internal energy
($\epsilon_0$) are chosen in this paper.
These parameters characterize the kinetic and thermal energy 
per particle in the injected outflow, respectively.
As $\epsilon_0$ and/or $\Gamma_0$ increases,
so does the kinetic and/or thermal energy per particle.
The rest mass density and pressure of
the injected outflow can be derived from these two parameters,
assuming an equation of state.
The larger $\epsilon_0$ and/or $\Gamma_0$ is, 
the smaller becomes the rest mass density.
For example, the rest mass density ($\rho_0$) is given as
\begin{eqnarray}
\label{density_1}
\rho_0 c^2 =
 \dot{E}_{0} [((1+\epsilon_0/c^2 \gamma){\Gamma_0}^2-\Gamma_0)v_0]^{-1}
    [\pi R_{0}^2]^{-1},
\end{eqnarray}
where $\gamma$ is adiabatic index.
The ideal gas equation of state, $p=(\gamma -1)\rho \epsilon$,
is employed,
where $p$ is pressure,
and the definition of specific enthalpy
$h/c^2(\equiv 1+\epsilon/c^2+p/\rho c^2)$ is used.
The equation (\ref{density_1}) is derived from
$T^{01}-\rho_0\Gamma_0 v_0=\dot{E}/(\pi R_0^2)$.

\begin{figure}
\epsscale{.80}
\plotone{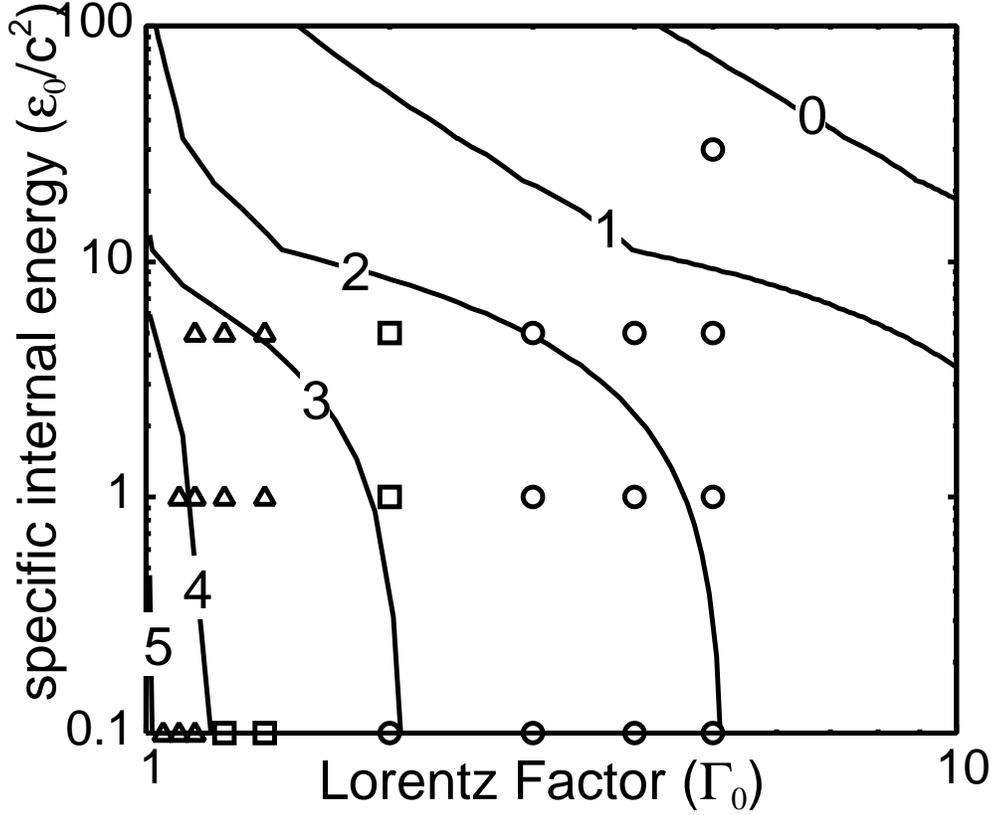}
\caption{The contours of the log-scaled mass density (in ${\mbox {g cm}^{-3}}$)
on the ($\Gamma_0$, $\epsilon_0 /c^2)$-plane.
The total energy flux is fixed to be $\dot{E}_0=10^{51}\mbox{ergs s}^{-1}$
and the radius of the injected outflow is $R_0=7\times 10^7$ cm.
The circles, squares, and triangles correspond to models,
in which the half opening angle of the outflow
at the time of breakout or at $t=10$ s is 
$\theta_{\rm sim} <5^{\circ}$, $5^{\circ}<\theta_{\rm sim} <20^{\circ}$, 
and $\theta_{\rm sim} > 20^{\circ}$, respectively.}
\label{epsilon_gamma}
\end{figure}

Figure \ref{epsilon_gamma} shows rest mass density
in the ($\Gamma_0, \epsilon_0$) plane
for fixed total energy, $\dot{E}_0=10^{51}\mbox{ergs s}^{-1}$
and the radius, $R_0=7\times 10^{7}\mbox{cm}$, of the
injected outflow.
The symbols in the plane present the calculated models.

Assuming that all the initial total energy is efficiently
converted to kinetic energy,
we can estimate the maximum bulk Lorentz factor from
the energy conservation law, 
$E_{tot}\approx M_{b}c^2\Gamma_{\rm max}$ \citep{Piran99}
where $M_{b}$ is baryon mass of the outflow.
Approximately the total energy is 
the sum of the rest mass, kinetic, and thermal energy,
$E_{\rm tot}\sim M_{b}c^2[1+(\Gamma_0-1)+\Gamma_0\epsilon_0/c^2]$.
Then the maximum Lorentz factor of a fire-ball can be estimated
as (Piran 1999)
\begin{equation}
\label{maxgam}
\Gamma_{\rm max}\sim\Gamma_{0}(1+\epsilon_0/c^2),
\end{equation}
which means almost all thermal energy is converted to kinetic energy
during expansion like SN explosion \citep{Arnett96}.

We will systematically vary $\Gamma_0$ and $\epsilon_0$
to survey any possible physical situations,
since we are unaware which is the case.
The condition which determines these parameters
should depend on the formation mechanism of 
the outflow from the center of the progenitor,
i.e., we should know in detail how the mass and 
the angular momentum are distributed in the progenitor,
how a massive star collapses,
how the neutrino transport occurs,
and how the magnetic field grows and affects the dynamics.
For example, a rapidly rotating massive star can
form geometrically thick accretion disk around
a new-born black hole or proto-neutron star
after the core-collapse.
The amount of the neutrino and anti-neutrino emission from 
such an accretion disk could be large.
Then the annihilation rate of neutrino and anti-neutrino
will be large enough 
so that thermal dominated outflow may be generated.
If the accretion disk is thin
due to a lack of rapid rotation of the progenitor,
conversely, the annihilation rate will be small,
which may results in the formation of a baryon rich outflow.
Note, however, that these conclusions may not be firm,
since we still do not know
the formation mechanism of this outflow.
Magnetic fields could be a key here.

Since there is a possibility that a non-relativistic but collimated
outflow is generated from the center of the progenitor,
we explore both the relativistic and non-relativistic outflows.
The bulk Lorentz factor, $\Gamma_0$, is varied from the relativistic
to the non-relativistic regimes,
$\Gamma_0=5, 4, 3, 2, 1.4, 1.25, 1.15, 1.1,$ and $1.05$ 
(which correspond to $v_0/c \sim$ 0.98, 0.97, 0.94, 0.87,
0.7, 0.6, 0.5, 0.4, and 0.3, respectively.).
The specific internal energy, $\epsilon_0/c^2$,
is changed from 30, 5, 1.0, and 0.5.
As a result,
the rest mass density of the outflow
is in the range from $2.4 {\mbox{g cm}}^{-3}$
[the fastest and hottest outflow : $(\Gamma_0,\epsilon_0/c^2)=(5.0,30)$]
to $4\times 10^4{\mbox{g cm}}^{-3}$
[the slowest and coldest outflow : $(v_0/c,\epsilon_0/c^2)=(0.3,0.5)$].
The pressure of the outflow also changes.
Table \ref{conditions} summarizes 25 calculated models.

To compare with \citet{Zhang03} we also list
the ratio of the total energy to kinetic energy, $f_0$,
is presented in the table,
where the rest mass energy is excluded from the total energy;
\begin{equation}
f_0\equiv {\rho_0\Gamma_0 (\Gamma_0-1)\over 
         \rho_0 (h_0/c^2) {\Gamma_0}^2-(p_0/c^2)-\rho_0 \Gamma_0},
\end{equation}
(see \cite{Zhang03}).
Because the definition of $f_0$ in  \cite{Zhang04}
was changed to be the ratio of kinetic energy to total energy,
we also show the inverse of $f_0$ in the table.
Our model A300, ($\Gamma_0,\epsilon_0$)=(5,30) is
similar one studied by \citet{Zhang04,Umeda05}.
The theoretically estimated Lorentz factor $\Gamma_{\rm max}$ from
Eq. (\ref{maxgam}) is also shown in Table \ref{conditions}.
Some cases, for example $(v_0/c,\epsilon_0/c^2)=(0.3,5.0$),
produce subsonic flow
and are, hence, difficult to calculate.
We exclude such cases from Table \ref{conditions}
to assure good numerical accuracy.
Since the mass density of the progenitor in the innermost region
for the computation is about $10^6 \mbox{g cm}^{-3}$,
all our models produce initially so-called light jet.
If the flow can keep the collimated structure in the progenitor,
such an outflow is expected to interact with the back flow 
\citep{Mizuta04} and has complex internal structures in the jet.
From eq. (\ref{maxgam}) we understand that
the most predominant case for the GRBs is the outflow with
very high $\Gamma_0$ and high $\epsilon_0$; i.e.,
model A300 ($\Gamma_{\rm max}\sim 150$).
The second one is model A50 ($\Gamma_{\rm max}\sim 30$).

\subsection{Hydrodynamic Equations}
\label{basicEQ}
We numerically solve two dimensional relativistic hydrodynamic equations,
assuming the axisymmetric geometry.
The basic equations are,
\begin{eqnarray}
\label{renzoku}
{\partial (\rho \Gamma) \over \partial t}+
{1\over r}
{\partial r(\rho \Gamma v_{r}) \over \partial r}+
{\partial (\rho \Gamma v_{z}) \over \partial z}=0,\\
\label{momen}
{\partial (\rho h\Gamma^2 v_{r}) \over \partial t}+
{1\over r}
{\partial r(\rho h\Gamma^2 v_{r}^2+p) \over \partial r}+
{\partial (\rho h\Gamma^2 v_{r}v_{z}) \over \partial z}={p\over r},\\
{\partial (\rho h\Gamma^2 v_{z}) \over \partial t}+
{1\over r}
{\partial r(\rho h\Gamma^2 v_{r}v_{z}) \over \partial r}+
{\partial (\rho h\Gamma^2 v_{z}^2+p) \over \partial z}=0,\\
\label{energy}
{\partial (\rho h\Gamma^2 -p) \over \partial t}+
{1\over r}
{\partial r(\rho h\Gamma^2 v_{r}) \over \partial r}+
{\partial (\rho h\Gamma^2 v_{z}) \over \partial z}=0,
\end{eqnarray}
$v_{i}$ ($i=r,z$) is the $i$-th velocity component.
The equations are written in the unit that
the speed of light is unity.
The updated version of
numerical hydrodynamic code used in \cite{Mizuta04} is
employed in this study.
The code adopts Godunov-type scheme which is advantageous for
capturing strong shocks with a few grid points in good accuracy.
In this version 
the physical values, such as, pressure, rest mass density
and the space components of 4-velocity, are used for this interpolation
using the Piecewise Parabolic Method (PPM) which allows us to get
third order accuracy in space.
The version of third order accuracy in space is used.
The results of 1D and 2D test calculations
are presented in the Appendix.

In this study, we calculate the propagation of the outflow 
crossing through the progenitor 
for 10 seconds after the outflow injection.
Since the outflow crossing timescale 
is shorter than the free-fall timescale,
we neglect the gravitational field by the core.
We also ignore self-gravity, for simplicity.
When an outflow propagates from the inner boundary 
towards the surface of the progenitor,
a strong bow shock appears and drives the progenitor gas
to high pressure and temperature.
Nucleosynthesis could occur in the such a hot and high density medium.
Since produced entropy due to the nucleosynthesis
is much smaller than that by the strong shock,
we do not consider nucleosynthesis as an energy source in the energy equation
(Eq. \ref{energy}).
We assume the ideal gas equation of state,
that is $p=(\gamma-1)\rho\epsilon$,
where $\gamma(=4/3 \ \mbox{constant in this study}$) is adiabatic index.

We used non-uniform grid points.
Logarithmically uniform 500 grid points are spaced for
$2\times 10^8 \mbox{cm} <z< 6.6\times 10^{10}  \mbox{cm}$.
We also set
uniform 120 zones for $0<r<1.2\times 10^9$ cm and
logarithmically uniform 130 zones for 
$1.2\times 10^9<r<1.1\times 10^{10}$ cm.
Some of our results, especially those having slower injection velocity
($v_0 \lesssim 0.7c$), exhibit spreading outflow, which has a large transverse
component of the flow with respect to the direction of the injected flow.
We find in the test calculations of 1D shock tube problem with
relativistic transverse velocity (see, \ref{SHtranseverse})
that the limited resolution of our numerical codes
produces some numerical errors
(numerical errors in both positional and
absolute values of density, pressure, 3-velocity,
up to several tens of percent).
However, in some slower models, such as, models G01, H01, and I01,
the flow is close to non-relativistic flows and
the errors caused by the relativistic transverse velocity are negligible.
In some faster models, such as, models A300, A50, A10, A05, etc,
the relativistic flow in mainly appear in the collimated jet
and propagates along the cylindrical axis.
Since the transverse velocity is not so large,
the numerical errors caused by the transverse velocity
are also negligible.
We can thus conclude that
the errors in mildly relativistic and spreading models
may affect to the opening angle,
but does not alter our main conclusion of a smooth transition
from the collimated jet to the expanding outflow
that is discussed in Sec. \ref{results}.
The achieved maximum Lorentz factors in each simulations
are not affected by this numerical error.

The boundary condition at $z=2\times 10^8$ cm is reflective except
the inner 7 grid points which are used to inlet outflow.
The reflective boundary condition is also employed at cylindrical axis.
The outflow boundary condition is employed at $r=1.2\times 10^9$ cm
and $z=6.6\times 10^{10}$ cm. 
We compellingly replace the numerical flux at the boundary
using injection conditions, such as,
$\rho_0$, $\epsilon_0$, and $\Gamma_0$, to inlet the outflow.
The numerical fluxes of the
mass, momentum for $z$ direction, and energy
at the injection points ($z=z_{\rm low}, 0<r<R_0$)
are given as $\rho_0 \Gamma_0 v_0$, $\rho_0 h_0 {\Gamma_0}^2 {v_0}^2 +p_0$,
and $\rho_0 h_0 {\Gamma_0}^2 v_0$, respectively.

\section{COLLIMATED JETS AND EXPANDING OUTFLOW}
\label{results}
\subsection{Collimated Jets: Cases with High $\Gamma_0$}
\label{collimated}
\begin{figure}
\epsscale{.60}
\plotone{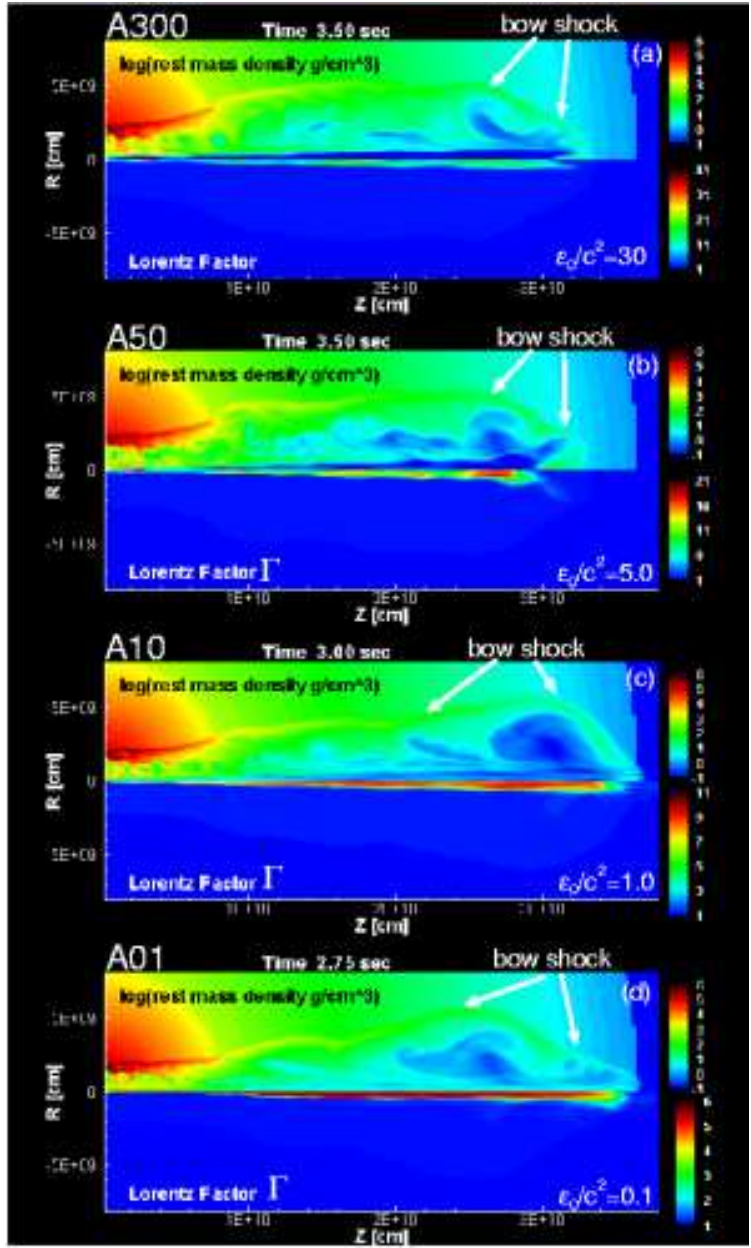}
\caption{The contours of the rest mass density (upper) and the
Lorentz factor (lower) of models,
from top to bottom,
A300, A50,, A10, and A01,
in which the Lorentz factor of injected outflow
$\Gamma_0$ is fixed to be 5,
at the time of
the breakout; 
$t=3.50$ s (A300), $t=3.50$ s (A50), $t=3.00$ s (A10),
and $t=2.75$ s (A01), respectively.
All models keep collimated structure.
A back flow, which runs from the head of the jet
in the anti-parallel direction, appears in each model.}
\label{panels}
\end{figure}

First, we show the results of the high Lorentz-factor cases 
with $\Gamma_0=5$; i.e., models A01, A10, A50, and A300.
Figure \ref{panels} shows the contours of rest mass density (upper)
and Lorentz factor (lower) in each model at the time of
breakout of the jet; $t=3.50$ s (A300),
$t=3.50$ s (A50), $t=3.00$ s (A30), and $t=2.75$ s (A01), respectively.
Note that the lower-left corner of each panel
does not correspond to the center of the progenitor
(see again Fig. \ref{condition_fig}).
The region with a large Lorentz factor is localized 
and is found only along the cylindrical axis.
This means that the outflow propagates throughout the progenitor, 
keeping a very collimated structure.
Thus we call this collimated outflow a jet.
The half opening angle ($\theta_{\rm sim}$) of the jet in each model
is kept small; it is only $\theta_{\rm sim} \sim 1^\circ$,
when the breakout occurs (see, Table \ref{conditions}).

The collimated jet is surrounded by the
shocked progenitor gas and back flow.
The width of the jet is about $10^9$ cm when the jet breaks.
Lateral shocks are found at distances of about
 $5-7\times 10^9$ cm from the $z$-axis.
Since the mass density of the injected outflow is much smaller than
that of progenitor near the center (see the previous section),
the velocity of the head of the jet
is less than that of the jet ($\sim c$),
as was analyzed by \citet{Norman83} and \citet{Marti97}.
The jet takes about 3 sec to cross over the progenitor.

The collimated jet eventually breaks out of the surface of the progenitor.
We observed small differences in the crossing times of the
jet within the progenitor among these models.
The jet in model A01 passes through the progenitor 
in a shorter time than that in model A10,
and, the jet in model A10 passes in an even shorter time
than that in model A50 and A300.  This can be understood,
since a lighter jet, i.e., lower density jet, feels larger 
resistance to proceed in the progenitor than a denser jet.

\begin{figure}
\epsscale{.80}
\plotone{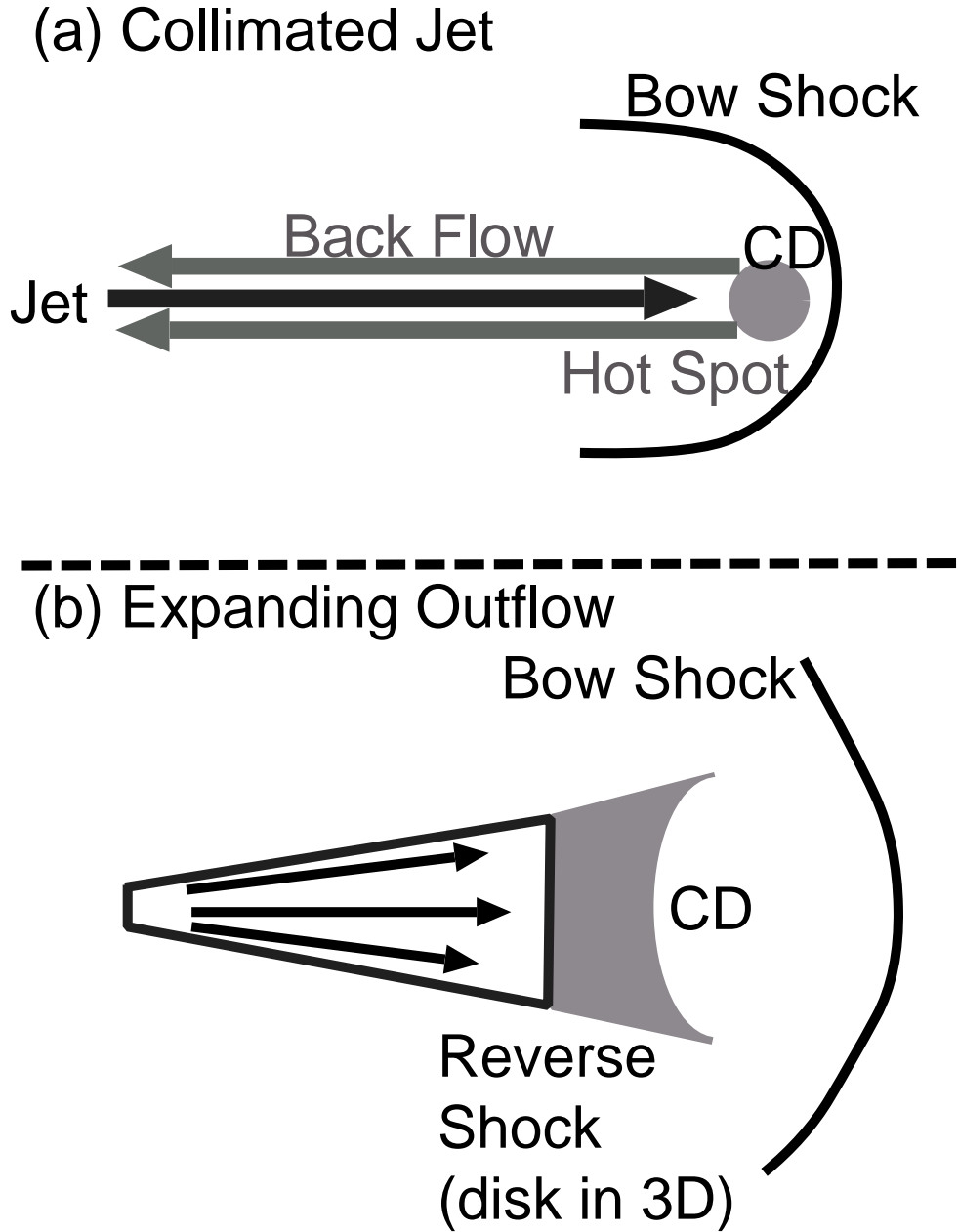}
\caption{Schematic figures of the collimated flows (a)
and the expanding outflow (b).
Only the main flow (outflow and back flow) is displayed here.
The back flow is usually meandering.
The half opening angle $\theta_{\rm sim}$ which is measured when the jet breaks out
of the progenitor is presented.}
\label{backflow}
\end{figure}

The dynamics and morphology of the calculated jet, such as its
collimated structure and the appearance of a back flow, 
are very similar to those shown by Zhang et al. (2003, 2004).
The flow structure is also very similar to those simulated
in the context of AGN jet propagation,
although most previous simulations have been done under the 
assumption of a constant ambient density.
One of the most prominent features of the jets in high-$\Gamma_0$
models is the appearance of the back flow
(see Fig. \ref{backflow}a).
In this figure, only main flow, i.e., a jet, and
back flow from the head of the jet are shown.

\begin{figure}
\epsscale{1.0}
\rotatebox{270}
{\plotone{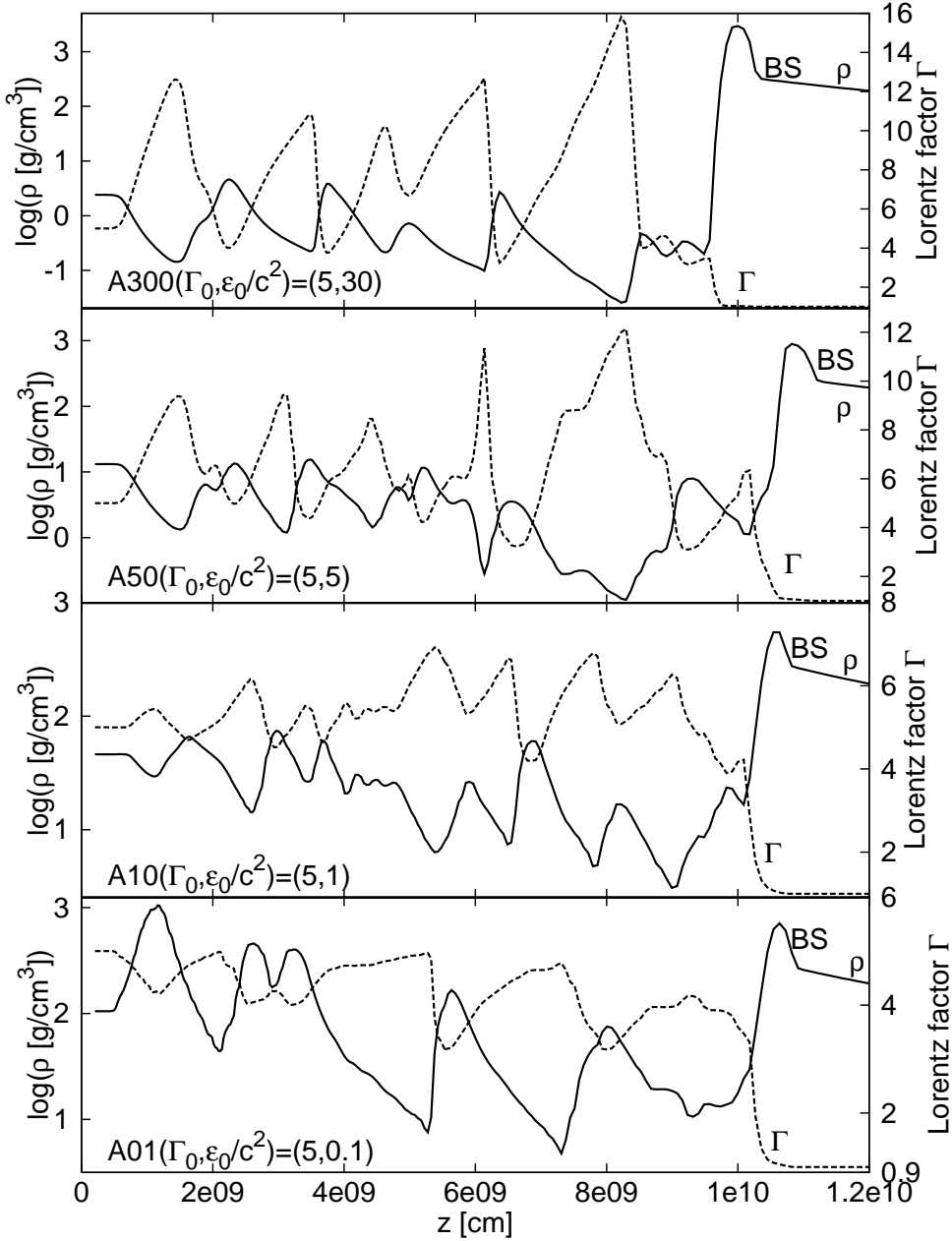}}
\caption{One dimensional profiles of density (solid line) and Lorentz factor
(dashed line) along the $z$ axis in the early phase of each simulation.
Several discontinuities can be seen.
Those correspond to the oblique shocks in the jet.
For the hotter outflow, Lorentz factor increases
at the oblique shocks.
\label{1Dpressure}}
\end{figure}

Since the jet propagates in the progenitor whose gas density
decreases outward, the hot spot does not clearly appear,
to the contrary to the simulation of the jet propagation into
a dense gas cloud with no density gradient.
At the head of the jet three discontinuities exist:
bow shock or forward shock (FS) which drives progenitor gas
to a high pressure and high temperature state,
contact discontinuity (CD), and
reverse shock (RS) or terminal Mach shock.
The kinetic energy of the outflow is converted to thermal energy
thorough the reverse shock.
The gas which passed this reverse shock forms a back flow
in the anti-parallel direction to that of the jet.
The path of the back flow is not straight but is meandering. 
The interaction between the main jet flow and the back flow 
enhances the oblique shocks in the jet flow.
As a result, the profile of the physical quantities,
such as density, pressure, and Lorentz factor,
is not monotonically decreasing nor increasing outward.
Figure \ref{1Dpressure} shows one dimensional profile (along the $z$ axis)
of the rest mass density and Lorentz factor of models
A300, A50, A10, and A01 in the early phase of the simulation.
The radial pressure profiles exhibit frequent up and down,
like a teeth of a saw, so does the density profiles
except at around the head of the flow.
Some discontinuities are seen along the $z$-axis,
although these features are difficult to see 
in the contours in Fig. \ref{panels}.
In model A10, A50, and A300 the Lorentz factor also shows 
a teeth-of-saw structure.
Those discontinuities correspond to the internal oblique shocks.

\begin{figure}
\epsscale{.35}
\rotatebox{270}
{\plotone{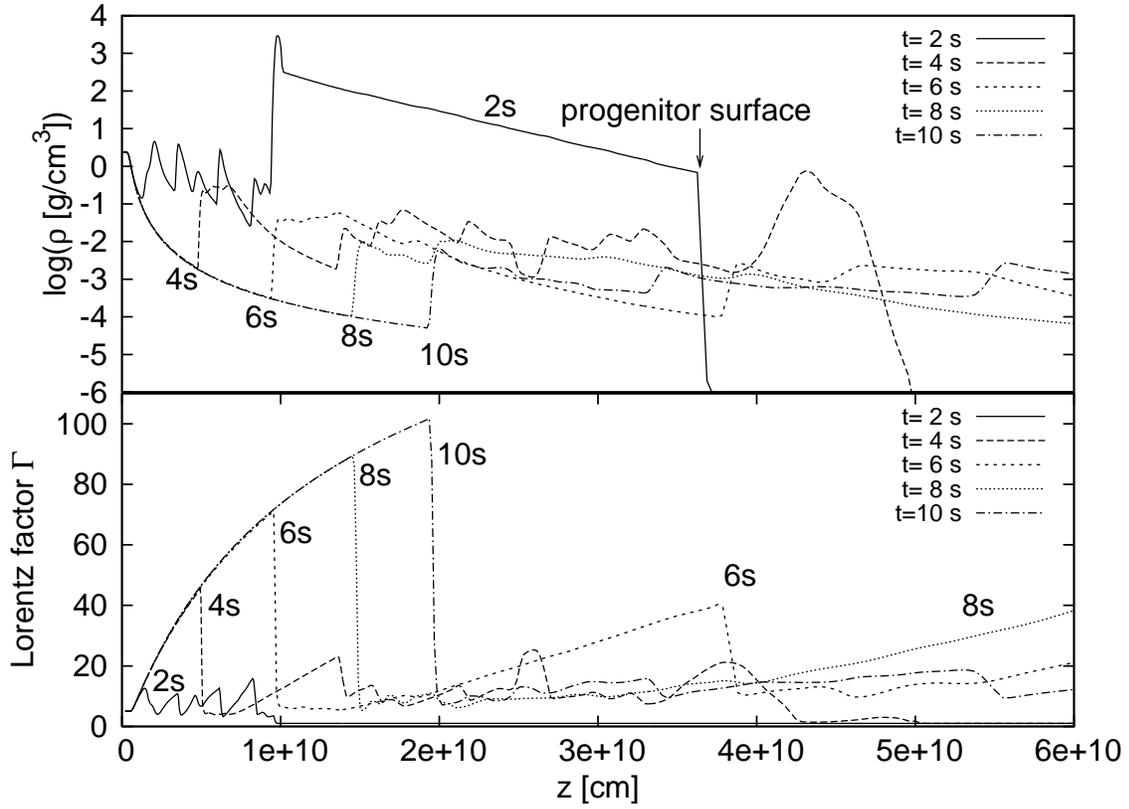}}
\caption{One dimensional profiles of density (top) and Lorentz factor
(bottom) along the $z$ axis of model A300
at different time ($t=$ 2, 4, 6, 8 and 10).
Maximum Lorentz factor is a few tens during propagation in the
 progenitor and increases larger than 100 after breakout from the progenitor.
\label{1Dseries}}
\end{figure}

Before the breakout the bulk Lorentz factor
increases during the propagation in
the progenitor only up to about 
43(A300), 20 (A50), 9.6 (A10), and 5.6 (A01),
respectively.
After the breakout, the bulk Lorentz factor further grows up
41 in model A50
and 104 in model A300
(see Fig. \ref{1Dseries} for a series of
one dimensional profiles of the rest mass density
and the Lorentz factor along the cylindrical axis of model A300).
The high Lorentz factor ($\Gamma \gtrsim 100$)
originates from the components which are injected in the later phase
($t\gtrsim 7$ s),
although the head of the outflow has already passed through
the boundary ($z=6.6\times 10^{10}$ cm).
This late appearance of the high Lorentz factor component is consistent
with the results by \citet{Zhang03,Umeda05}.
The maximum Lorentz factor in each model is in good agreement with
the ones predicted by Eq. (\ref{maxgam}),
(see, table \ref{conditions}
in which the maximum Lorentz factor ($\Gamma_{\rm max,sim}$)
achieved in the calculation and theoretically estimated value
$\Gamma_{\rm max}$ are presented.).
Note that in model A300 the maximum Lorentz factor is expected to
achieved after $t = 10$ s and is thus not shown in Fig. \ref{1Dseries}.
When an outflow is injected to the computational domain,
the radius of the outflow increases via adiabatic expansion.
At this time the thermal energy is converted to
the kinetic one and thus the Lorentz factor increases.
This implies that the acceleration occurs
due to the effective conversion of the thermal energy
to kinetic energy during the propagation in the progenitor.
As we show in Fig.\ref{1Dpressure},
the Lorentz factor does not monotonically increase upward
but decreases at the points where the density increases.
These points correspond to the locations of oblique shocks.  
There exist not only oblique shocks but also rarefactions, in
which the density decreases while the Lorentz factor increases.

\begin{figure}
\epsscale{.8}
\plotone{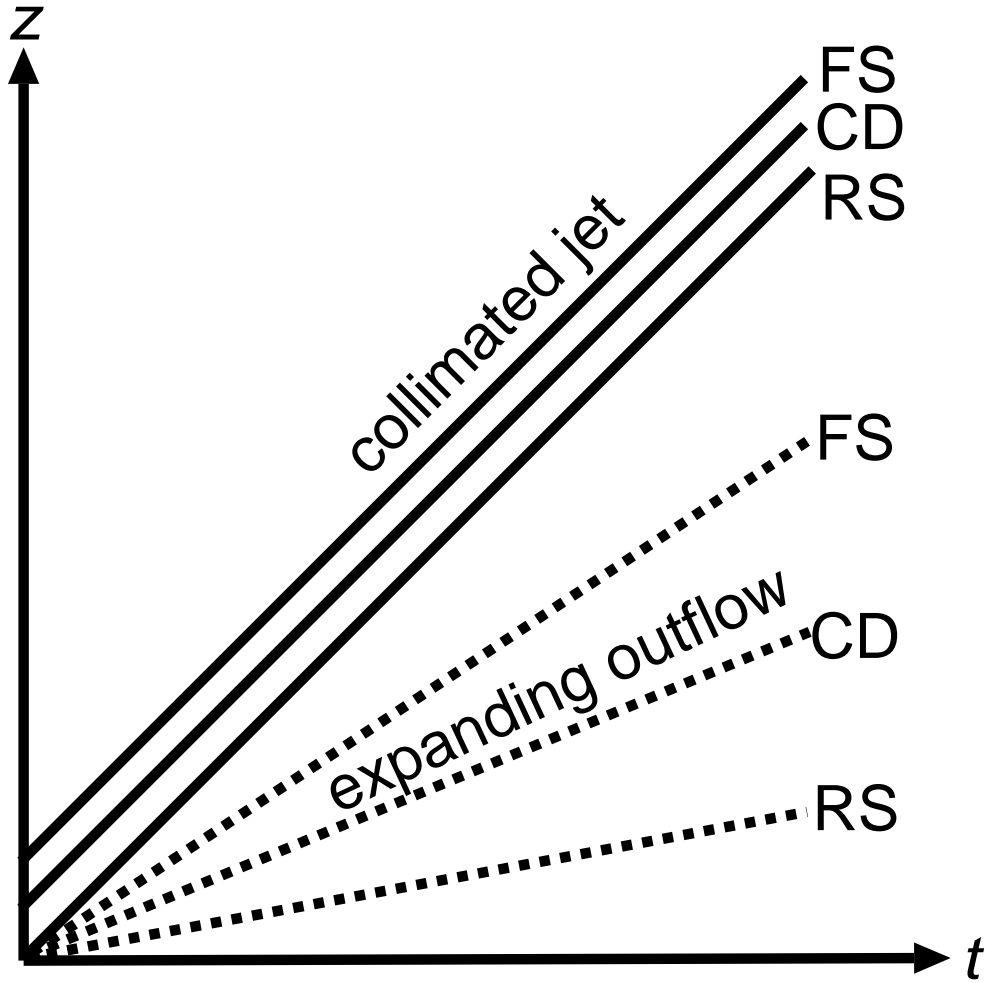}
\caption{A figure of the positions of froward shock, contact discontinuity,
and reverses shock for cases of collimated jet (solid line)
and expanding outflow (dashed line). 
\label{foward_reverse}}
\end{figure}

The pressure inside of the bow shock is very smooth
except at the places where discontinuities appear.
These discontinuities follow each other,
i.e. the separations between them do not grow with time
(see, Fig \ref{foward_reverse} which presents schematic figure of
the positions of discontinuities).
This feature is quite different from the shocks produced 
by a spherical explosion, such as supernova remnant,
where the FS and RS separate with time.
Most of the kinetic energy is converted to thermal energy
at the RS.  The hot gas between the RS and CD is exhausted 
to lateral direction and finally generates a back flow.

\begin{figure}
\plotone{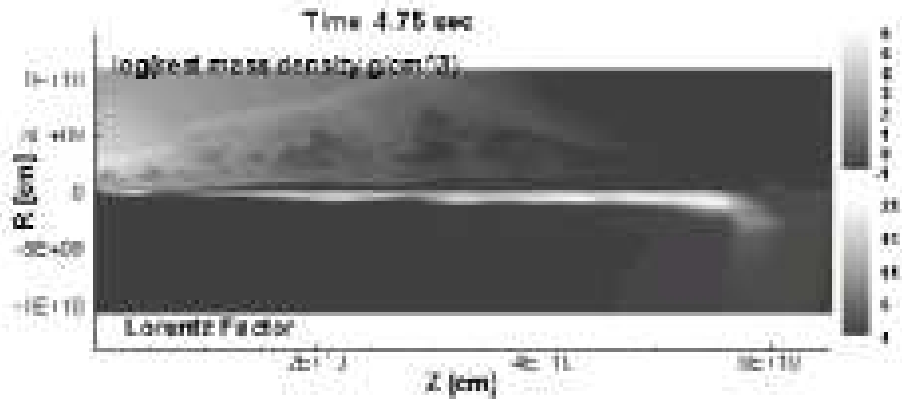}
\caption{Same as Fig. \ref{panels} but for model A50 at $t=4.75$ s.
All the flow components, including the back flow
during the propagation in the progenitor, become outflow. 
The region with high velocity gas remains along the $z$-axis.
\label{eruption}}
\end{figure}

Even after the breakout of the outflow from the progenitor surface
we continued to calculate the propagation of the jet up to
the distance of $6.5 \times 10^{10}$ cm.  The outflow 
exhibits a significant expansion after the breakout.
Fig. \ref{eruption} shows rest mass density (upper) and
Lorentz factor (lower) of model A50 at $t=4.75$ s, which 
illustrates how the jet breaks out and expands to the ISM.
Although the shocked progenitor gas and the cocoon which
comprises the back flow are confined within a narrow zone
by a strong bow shock before the breakout, 
this is no longer the case after the breakout.  In fact,
all the progenitor gas, including the back flow and the
cocoon, begin to expand and forms a global outflow towards the ISM.
Note, however, that still the high velocity 
component survives along the $z$-axis.
As a result, high velocity component is
surrounded by slow velocity and dense component.
We will further follow jet propagation in near future.

\subsection{Expanding Outflows : Cases with Low $\Gamma_0$}
The outflows with smaller injection velocity behave very differently,
compared with those with larger injection velocity presented above.
In this section we discuss models G01, G10, and G50, in which 
the injection velocity is fixed to be $v_0=0.5c$ ($\Gamma_0=1.15$).

\begin{figure}
\epsscale{.80}
\plotone{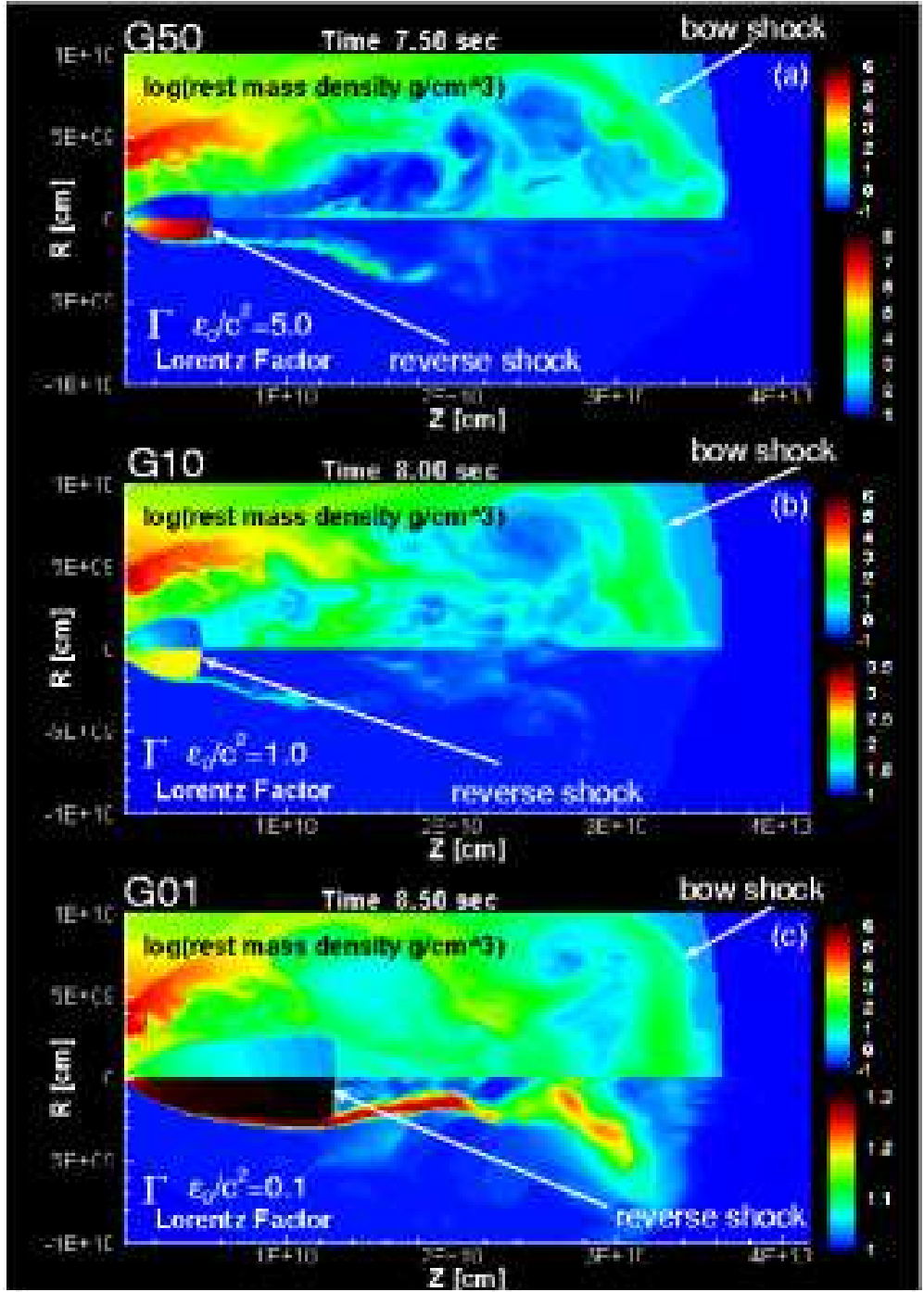}
\caption{Same as Fig. \ref{panels} but for models
G50 (top), G10 (middle), and G01 (bottom),
in which the velocity of injected outflow
$v$ is fixed to be $0.5c$, at $t=7.5$ s (G50),
$t=8.0\mbox{ s}$ (G10), and $t=8.5\mbox{ s}$ (G01), respectively.
The outflow has an expanding structure like a fan.
As time goes on, the distance between the reverse shock
(RS) and the forward  shock (FS) increases, just
as in a supernova remnant.} \label{panels2}
\end{figure}

Figure \ref{panels2} shows the contours of rest mass density 
and Lorentz factor of these three models at $t=7.5\mbox{ s}$ (G50),
$t=8.0\mbox{ s}$ (G10), and $t=8.5\mbox{ s}$ (G01), respectively.
In contrast with the previous cases
the outflow shows an expanding structure like a fan.
Let us remind that the injected outflow is initially parallel 
to the cylindrical axis.
This expanding structure can be seen from the early phase of the
evolution of the outflow.  
We also see that the RS separates from the FS with time in these cases.
The high velocity region can be seen around the
innermost region which ends at the reverse shock,
forming a ``disk''-like region with a large cross section
(hereafter called as a disk),
where the cross section stands for the area
of the disk in the three dimensional space
(see figure \ref{backflow}b and \ref{foward_reverse}).
Interestingly, the separation between
the FS and RS increases with the time in the present cases,
just as in supernovae remnant or pulsar wind, which makes 
a good contrast with the cases of the collimated jet.
As the outflow expands with a large opening angle, 
the cross section of the CD increases.  Since the cross section 
is proportional to square of the radius of the outflow,
the flow with a large opening angle needs so large 
dragging power to proceed.
This suppresses the back flow formation at the head of the jet.
As the cross section of the CD increases,
the gas is gradually collected at the head of the flow
because the FS sweeps up the progenitor gas.
This works to enhance the expansion of the outflow to the lateral
direction, thus an opening angle being increased.
The half opening angle $\theta_{\rm sim}$ of the outflow at the break is 
22$^\circ$ (G01), 30$^\circ$ (G10), and 26$^\circ$ (G50), respectively.

\subsection{Intermediate Cases}
\begin{landscape}
\begin{figure}
\hspace{-3.5cm}
\epsscale{1.0}
\plotone{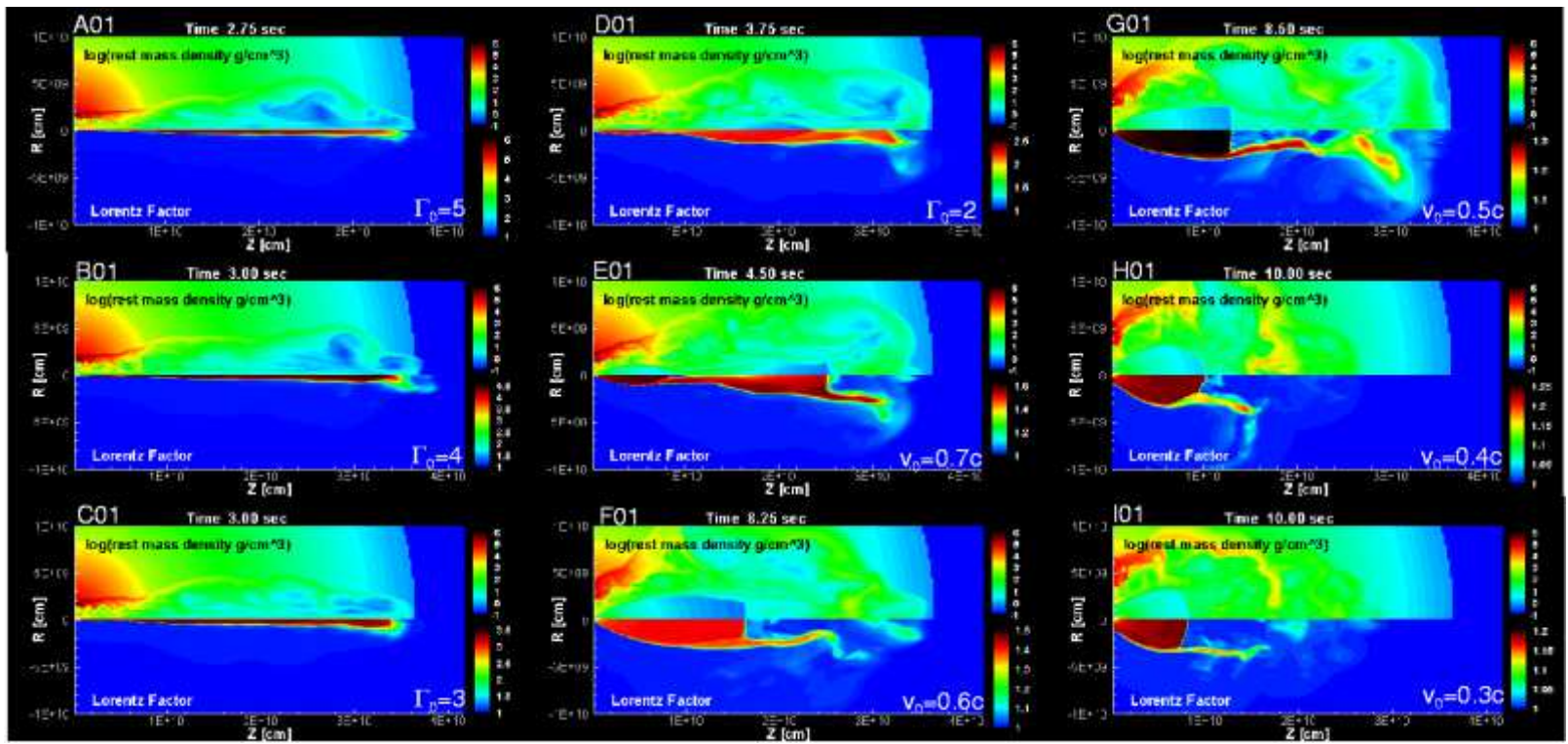}
\caption{Same as Fig. \ref{panels} but for models
A01, B01, C01, D01, E01, F01, G01 H01, and I01.
The transition in the outflow dynamics from the collimated structure
to the expanding structure is evident.
} \label{panels3}
\end{figure}
\end{landscape}

The continuous transition from the collimated jet
to the expanding outflow takes place, as the Lorentz factor
(and thus velocity) of the injected outflow decreases.
This is nicely illustrated in Figure \ref{panels3},
which shows the density and $\Gamma$ contours of models with
a variety of $\Gamma_0$ values.
The specific internal energy of the injected outflow
is fixed to be $\epsilon_0/c^2=0.1$.
For models A01, B01, C01, and D01 we find a collimated jet,
the half opening angle of which is only a few degrees.
The outflow starts to show an expanding tendency from model
 E01 $(v_0/c=0.7,\epsilon_0/c^2=0.1)$,
in which the half opening angle $\theta_{\rm sim} \sim 12^\circ$
at the time when the outflow breaks out,
and finally model I01 exhibits a typical expanding-fan structure
with $\theta \sim 32^\circ$ at $t=10$ s (see Table \ref{conditions}).
The radius of the outflow gradually increases as the velocity of
injected outflow decreases.
The sideway expansion of the bow shock becomes significant
when $v_0$ is small, $v_0/c < 0.7$.

The same transition also takes place in other series of
models.  We next fixed the internal energy to be
$\epsilon_0/c^2=1$ and $\epsilon_0/c^2=5$.
The opening angle in each model varies
from a few degrees (model A10, and A50)
to more than $30$ degrees (models G50 and H10).
The transition takes place in models D10 and D50, in which
the specific internal energy $\epsilon_0/c^2$ if fixed to
1.0 and 5.0, respectively.
We summarize these results in Fig. \ref{epsilon_gamma},
in which circles, squares, and triangles indicate the cases
with a half opening angle of the outflow
at the time of the breakout or at the time $t=10$ s being
$\theta_{\rm sim} <5^\circ$, $5<\theta_{\rm sim} <20^\circ$, and 
$\theta_{\rm sim} > 20^\circ$, respectively.
The RS separates from the CD and the FS 
in cases of the expanding outflow.
The maximum Lorentz factor $\Gamma_{\rm max,sim}$ achieved 
in the calculation
in every model is presented in Table \ref{conditions}.
Those are good agreement with theoretical estimated $\Gamma_{\rm max}$.
The numerical errors caused by the relativistic transverse velocity
respect to the propagation direction described
in the \ref{basicEQ} and  \ref{shtubetrans} are only appreciable in the
intermediate cases between collimated jet to expanding outflow.
We estimate that the numerical errors produce an
uncertainly of 10 percent in the Lorentz factor
at which the transition happens.
Our main conclusion is not affected by the numerical errors
of this sort, however, since the numerical errors are negligible
in outflow which keeps quite good collimation, as well as in
non-relativistic outflows.

\subsection{Dependence on the Energy Injection Point}
We do not know where and how the outflow forms and evolves
in the progenitor at present.
To see how sensitive the jet propagation properties are to
the injection point, we also calculated model A50b,
in which the lower boundary is set to be 
$2\times 10^7$ cm from the center; that is,
$1.8\times 10^8$ cm closer than in other cases ($2\times 10^8$ cm).
Other model parameters are the same as those of model A50.
Since the progenitor has a very large density gradient 
in the radial direction at small radii,
it is not trivial weather the outflow can drill this ``wall''
in this case.  Note that \citet{Aloy00} also set 
the inner boundary at $2\times 10^7$ cm
and deposited thermal energy around the boundary.

\begin{figure}
\epsscale{.8}
\plotone{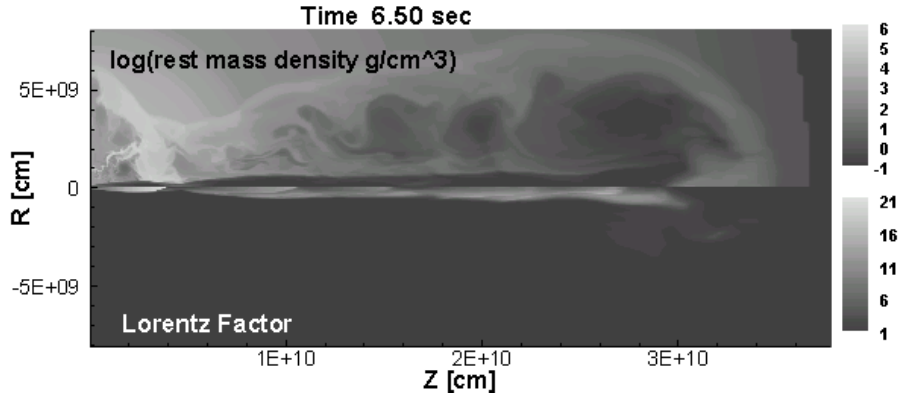}
\caption{Same as rest Fig. \ref{panels} but for model A50b
in which the inner boundary is set to be at $2\times 10^7$ cm from the center.
In the beginning of the evolution it takes slightly a longer time
to drill high density region.
After that the dynamics and morphology are very similar to
those of the regular case (A50).}
\label{inner}
\end{figure}

Fig. \ref{inner} shows the results of model A50b
at the time $t=6.5$ s when the outflow breaks out. 
There is a difference in the early phase ($\lesssim 3$ s) of the dynamics
compared with model A50.
The higher density of progenitor gas prevents
the outflow from proceeding to the direction of $z$ axis.
The outflow gradually drills the progenitor, and 
the bow shock spreads to lateral direction.
Except for these subtle differences,
the morphology and dynamics are very similar,
once the outflow drills out the high density region.
The outflow can keep the collimated structure
during the propagation in the progenitor.
A back flow appears in both cases.
The outflow can finally break out the progenitor
as shown in case A50.

\section{DISCUSSION}
\label{discussion}
\subsection{Physics discriminating two types of outflow}
We observe a dramatic but smooth transition from collimated 
structure to expanding one in the outflow propagation 
by changing the Lorentz factor and specific internal energy of
the injected outflow.  
What is the key physics which is responsible for the transition ?
{One of them is the pressure of the injected outflow.
The pressure of the injected outflow in the models
of the collimated jet is lower than that of expanding outflow.
Additionally,} we, here, point out that the appearance of
the (internal) reconfinement shock could be a key,
since the lateral expansion is suppressed by the presence of
the reconfinement shock \citep{Mizuta04}, thus
leading to a formation of the collimated jet structure.
In fact, high-pressure regions driven by a bow shock
can keep the collimated structure.
Models which have expanding structure have
no or less internal structure in the reverse and side shocks.
Then the injected flow is bi-forked.
On the contrary, models which keep the collimated structure
have a number of internal oblique shocks within the jet.
The existence of a back flow also enhances the appearance of such shocks.
Such shocks may allow the magnetic field generation in the jet
(discussed later).

The internal shock model was introduced by 
Rees \& Meszaros (1994 ; see also Kobayashi et al. 1997, and
 Spada et al. 2001)
to explain very short time variation of GRBs.
It is predicted that the outflow has
up to a few hundred internal structures or ``shells''
and multiple two shell collisions
occur at the distance of $10^{13-15}$ cm
from the fire ball which emits gamma-rays.

A simple linear analysis by \citet{Urpin02} and \citet{Aloy02}
concludes that the timescale for a jet to propagate to 
the progenitor surface is long enough for perturbations to grow
to form internal shocks and variability in GRBs.
The instability is caused by the shear flow or perturbation in
pressure, density, and/or velocity in lateral direction in the jet.
The derived growth rate of the shear instability monotonically 
increases with the decrease of the wavelengths \citep{Urpin02}.
Although with our current resolution it is impossible to resolve
such fine structure, we can estimate the timescale of the instability,
assuming the specific internal structure, say, that seen in
Fig \ref{1Dpressure}.
Taking the wavelength to be a typical interval of the oblique shocks,
$\sim 10^9$ cm, and bulk Lorentz factor to be $\sim 5$,
we find that the timescale of the shortest mode in model A50 
is on the order of $10^{-2}$ s.
This growing timescale is sufficiently shorter than the dynamical one;
that is, there is ample time for perturbations to grow to shocks,
as was claimed by \citet{Aloy02}.

In addition to the shear instability scenario, there exists
another possible cause of the internal oblique shocks;
that is the interaction between the jet and the back flow.
As the path of the back flow is not straight,
the boundary between main jet and the back flow is not smooth but
dynamically perturbed.
Since the dynamics is so complicated and is non-linear,
it is hard to specify which mechanism, either the growth of the 
instability in the jet or the interaction between the jet and the back flow,
is dominant for the appearance of the internal shocks.
The detailed structure should depend on the resolution of the
calculation.

\begin{figure}
\epsscale{.8}
\plotone{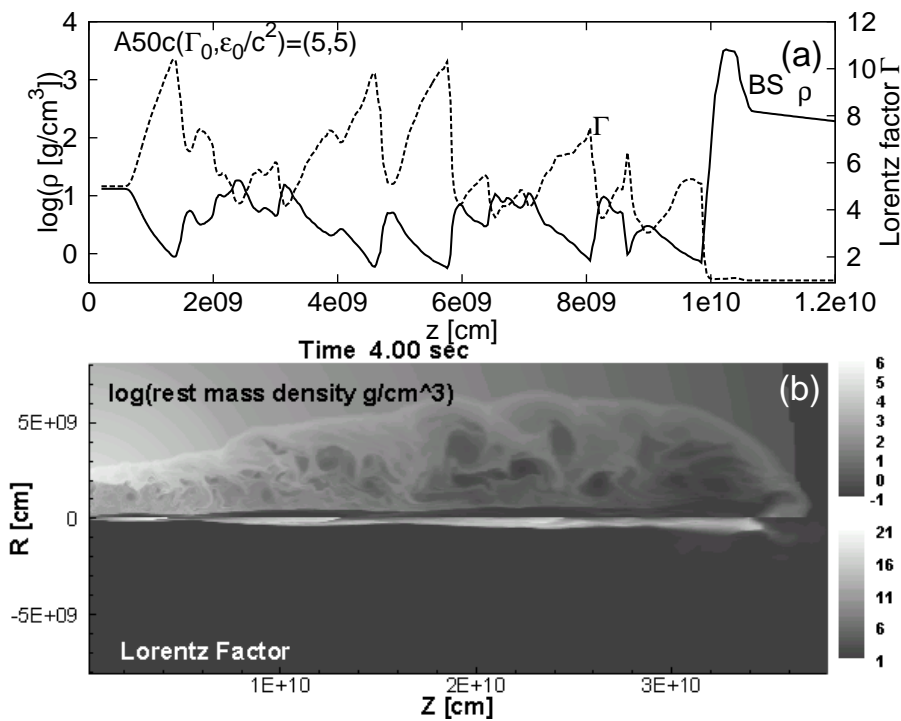}
\caption{(a) One dimensional rest mass density (solid line)
and Lorentz factor (dashed line)
profiles along the $z$-axis of model A50c, which is
twice higher resolution version of model A50.
Fine structure can be seen (see the result shown in
 the bottom of the Fig. \ref{1Dpressure} for comparison.).
(b) Same as Fig. \ref{panels} but for model A50c.
Complex structures can been seen in the cocoon caused by 
the mixing of back flow and shocked progenitor gas.}
\label{panels4}
\end{figure}

For this reason, we have calculated the same model as model A50
but with twice higher resolutions than that of model A50;
that is model A50c.
Fig. \ref{panels4}a shows the results of model A50c
(see the bottom panel of Fig. \ref{1Dpressure} for comparison).
The discontinuities still clearly appear in A50c, as well,
but the number of discontinuity has increased than that of model A50.
The cocoon has fine structures and vortices.
The emergence of vortices enhances
the mixing of the back flow and shocked progenitor gas.

The appearance of the oblique shocks in the jet could
be a key to the collimation.
Whether such internal shocks appear or not strongly 
depends on how high pressure can be achieved
by the presence of the bow shock.
But it is difficult to predict 
whether the outflow expands or keeps collimated structure
by a simple formula.
What is clearly demonstrated through our simulations is that
the large Lorentz factor and/or internal energy
of the outflow can make a strong bow shock ahead of the outflow,
leading to a good confinement of an exploding hot outflow.

Before closing this section, we wish to remark on the
role of magnetic fields.
The GRBs quite generally exhibit power-law and non-thermal 
spectra produced by synchrotron emission.
Then, we need moderately strong magnetic fields within the
jet, but it is not well understood how the magnetic field 
can be generated and grows in GRBs.
A plausible mechanisms for creating magnetic fields
is the Weibel instability which sets out
when the velocity field is not isotropic
and which can amplify magnetic fields quickly.  Further,
the particle acceleration to highly relativistic regime 
can take place in the shocks.
\cite{Nishikawa05} showed relativistic electrodynamics
particle simulations of launching jet into the ambient medium.
They observed the amplification of non-uniform and small-scale
magnetic fields by the Weibel instability.
Although they studied the shock at the head of the jet,
magnetic fields can also be amplified in internal shocks
within the jet.

If the magnetic field grows up, it will inevitably
affect the emissivity of
pre-cursors and emissivity of gamma-rays of GRBs
after breakout from the progenitor.
If large-scale magnetic fields can be created, 
they should affect the dynamics of the outflow.
We need further studies in this field.

\subsection{GRBs, X-ray Flashes, and SNe}
Finally we discuss the astrophysical implications of our results. 
We have shown different types of outflow dynamics
which can arise even by the input of the same
total energy power through the initial outflow.
The input outflows may propagate, keeping a collimated structure
in the progenitor, or showing an expanding structure.
The change of the outflow shape with changes in $\epsilon_0$
and $\Gamma_0$ is gradual (see Fig. \ref{epsilon_gamma}).
Although we cannot directly observe the propagation of the outflow
within the progenitor due to a large optical depth,
different dynamics will produce observable effects
after the outbreak of the outflow.
Different outflow dynamics may account for a variety of phenomena.
The outflow with significant collimation and high 
Lorentz factor ($> 100$) will produce GRBs.
When the Lorentz factor in the collimated jet is a bit smaller,
so is the peak energy of the emission.
Such an outflow could be observed as XRFs.

Some of our simulations show an expanding outflow even when
$\Gamma_0$ and/or $\epsilon_0$ are relatively small.
Such outflows 
may be observed as less energetic explosions as XRFs,
even if the viewing angle is relatively small.
These cases may correspond to the high velocity
but non-relativistic flow, such as SN2002ap,
since the flow still has a directivity.

There is another factor for explaining
the differences between GRBs and XRFs; 
that is the viewing angle to the jet.
Recently the unified model has been proposed by
\citet{Nakamura00} and by \citet{Yamazaki02,Yamazaki04}
who claim that the different phenomena are attributed to
different viewing angles, 
although the bursts themselves are identical.
If the viewing angle is less or near the jet opening angle,
the object will be observed as a GRB.
If not, it will be identified as an XRF.
\citet{Zhang04} derived the observational energy
as a function of the viewing angle based on their simulations.

The third but less unlikely possibility has been pointed out
through the discovery of Soft Gamma-ray repeaters (SGR) 1806-20.
Because of its proximity its spectral and light variations 
have been recorded in unprecedent details 
\citep{Terasawa05,Palmer05}.
Such an event may be observed as a short gamma-ray bursts,
if it explodes in nearby galaxies, although its spectrum is
basically blackbody and does not agree with those of GRBs.

Finally, we comment on the synthesis of heavy elements.
The large cross section of the forward bow shock
is advantageous for the synthesis of heavy elements, since 
then wider regions are available for the nucleosynthesis
(MacFadyen and Woosley 1999; Nagataki et al. 2003).
We find that
the outflows have an expanding structure in some cases.
This expansion is reminiscent of the cases
of aspherical explosions of supernova or hypernova as was
shown in Nagataki et al. 2003 (1998, see also
Nagataki 2000, Maeda \& Nomoto 2003, Nagataki et al. 2006).
A large amount of heavy elements is expected to be synthesized
by such an aspherical expansion, since the effective cross 
section of the bow shock is very large.
The difference of the expansion should affect
the amount of the synthesized elements.
The viewing angle is also an important parameter for this
type of outflow and is expected to explain the observed features 
of SNe \citep{Mazzali05}.

\citet{Nagataki03} calculated explosive nucleosynthesis by 
hydrodynamic calculations along the line of the collapsar model.
It is also possible that
the nucleosynthesis of heavy elements can occur in the
accretion disk surrounding the black hole
\citep{Woosley99,MacFadyen99,Fujimoto01,Pruet03}.
\citet{Nagataki03} showed that ${}^{56}$Ni is synthesized in the
jet like outflow and concluded that the amount of ${}^{56}$Ni
becomes larger when the energy deposition occurs in a short time,
since then the deposited energy is effectively converted 
to thermal energy by the strong shock.
Such calculations can be a good diagnostic tool 
to compare with the observations,
since the amount of synthesized elements can be estimated.

\section{CONCLUSION}
\label{conclude}
We investigate the propagation and dynamics of the outflows
in the progenitor in the context of collapsar model
for a central engine of GRBs by means of hydrodynamical simulations.
We assume a fixed power input of the initial outflow of
$\dot{E}_0=10^{51}\mbox{ergs s}^{-1}$ and the radius of the
injected outflow $R_0=7\times 10^7$ cm,
and follow the propagation
of hot outflow for different values of $\epsilon_0$ and $\Gamma_0$
over the ranges of $1.05 \le \Gamma_{0} \le 5 \ (0.3 \le v_0/c \le 0.98)$
and $0.1 \le \epsilon_0/c^2 \le 30$.
The net energy for 10-second injection satisfies the explosive energy of
so-called hypernovae $\sim 10^{52}$ ergs.

Our conclusions can be summarized as follows:
\begin{enumerate}
\item 
The propagation dynamics of the outflow dramatically changes
from the collimated structure to the expanding one 
as $\Gamma_0$ decreases.
If the Lorentz factor is high enough, say $\Gamma_0 > 3$,
the outflow can propagate throughout the progenitor,
keeping a very collimated structure.
The half opening angle is
$\theta_{\rm sim} < 2^\circ$ for $\Gamma_0 \gtrsim 3$.
But the opening angle has weak dependence on $\epsilon_0$, 
as well;
we get $\theta_{\rm sim} < 3^\circ$ even for smaller $\Gamma_0$ 
but with small $\epsilon_0/c^2 \lesssim 0.1$.
The maximum Lorentz factor is, on the other hand,
sensitive to both of $\Gamma_0$ and $\epsilon_0$;
roughly $\Gamma_{\rm max} \sim \Gamma_0 (1+\epsilon_0/c^2)$.

\item
In the relativistic, collimated flow,
a back flow, which is anti-parallel to the main jet, appears.
During the propagation in the progenitor,
we can see some internal structures caused
by the instability grown by the shear flow in the jet
or by the interaction between the jet and back flow
in the collimated jets.
Such oblique shocks can help the reconfinement of the jet.
The maximum Lorentz factor of the jet follows
a simple formula derived from energy conservation relation.
After the breakout the outflow expands into the interstellar space,
although there still remains a high velocity component 
along the $z$-axis.  Its half opening angle if a few degrees.
This could be observed as GRBs.
Another flow component which surrounds the central high velocity 
component can also be seen.
It originates from the back flow during the propagation in
the progenitor and shocked progenitor gas.

\item
When the Lorentz factor (or the initial velocity) of the 
outflow is not large, say, $\Gamma_0 \lesssim 1.4$ ($v_0/c \lesssim 0.7$),
the outflow no longer keeps the collimation and thus expands
to the forward and lateral directions.  
Eventually the outflow breaks out like an aspherical
supernova explosion.
This is because with the small outflow velocity the bow shock is
weak and cannot drive the progenitor gas to high enough pressure.
As a result, the reconfinement shocks, 
which are necessary for the collimation, does not appear.
Thus, the structure is relatively featureless in the outflow.
As the cross section of the reverse shock increases with time,
the mass is collected at the head of the outflow.
This enhances lateral expansion.

\item
High Lorentz factor ($>10$)is needed
to explain energetic phenomena, such as GRBs and XRFs,
but the different initial internal energy affects the opening
angle of the outflow for injected outflows
of smaller Lorentz factor, i.e. slower velocity, thereby producing
a marked difference in its observable.
Rather low internal energy, $\epsilon_0/c^2 \lesssim 0.1$,
and relatively small Lorentz factor ($<$ 5)
leads to collimated non-relativistic jets,
which will be observed as a failed GRB.
High internal energy, $\epsilon_0/c^2 \gtrsim 5$,
leads to un-collimated relativistic jets, which
could be observed as XRFs.
We can thus phenomenologically explain different
types of explosions, GRBs XRFs, and failed GRB along the same line
but with different values of $\epsilon_0$ for slower injected velocity.
However, a cause of producing a variety of $\epsilon_0$
and $\Gamma_0$ is still unknown.
It should be investigated in future work.
\end{enumerate}

We thank W. Zhang for his comments on the definition of $f_0$
in their paper.
We appreciate S. Woosley for his helpful comments to this
manuscript.
We gratefully acknowledge
M. A. Aloy, E. M\"uller, A. Macfadyen, N. Ohnish, and M. Horikoshi
for the discussion on the numerical method.
One of the authors (A.M.)
was supported by a Research Fellowship
of the Japan Society for the Promotion of Science.
This work was supported in part by the Grants-in-Aid of the Ministry
of Education, Science, Culture, and Sport  (14079205, A.M. and S.M),
(16340057 S.M.) and (14102004, 14079202, and 16740134, S.N.),
This work was supported by the Grant-in-Aid for the 21st Century COE
"Center for Diversity and Universality in Physics" from the Ministry of
Education, Culture, Sports, Science and Technology (MEXT) of Japan.

This work was carried out on NEC SX5, Cybermedia Center and
Institute of Laser Engineering, Osaka University,
SX8 at YITP in Kyoto University,
and Fujitsu VPP5000 of National Observatory of Japan.
We appreciate computational administrators for technical supports.

\begin{appendix}
\section{1D and 2D Test Calculations}
Since the relativistic hydrodynamic code used in this study
is updated version of the one used in \citet{Mizuta04}.
In this appendix, we describe the numerical methods employed in the code,
and show the results of 1D and 2D
numerical test problems that have been used by other groups
to show the ability of our new code.

Relativistic hydrodynamic equations can be written in 
conservative form as follows.

Cartesian coordinate:
\begin{eqnarray}
{\partial \mbox{\boldmath $u$}\over \partial t}+
{\partial \mbox{\boldmath $f$}(\mbox{\boldmath $u$})\over \partial x}+
{\partial \mbox{\boldmath $g$}(\mbox{\boldmath $u$})\over \partial y}=0,
\end{eqnarray}

Cylindrical coordinate:
\begin{eqnarray}
{\partial \mbox{\boldmath $u$}\over \partial t}+
{1\over r}
{\partial (r \mbox{\boldmath $f$}(\mbox{\boldmath $u$}))
\over \partial r}+
{\partial \mbox{\boldmath $g$}(\mbox{\boldmath $u$})\over \partial z}=
\mbox{\boldmath $s_c$}(\mbox{\boldmath $u$}),
\end{eqnarray}

Spherical coordinate:
\begin{eqnarray}
{\partial \mbox{\boldmath $u$}\over \partial t}+
{1\over r^2}
{\partial (r^2 \mbox{\boldmath $f$}(\mbox{\boldmath $u$}))\over \partial r}+
{1\over r \sin \theta}
{\partial (\sin \theta \mbox{\boldmath $g$}(\mbox{\boldmath $u$}))
\over \partial \theta}=
\mbox{\boldmath $s_s$}(\mbox{\boldmath $u$}),
\end{eqnarray}
where, {\boldmath $u$} is conservative vector,
{\boldmath $f$} and {\boldmath $g$} are flux vectors,
{\boldmath $s_c$} and {\boldmath $s_s$} are source vectors,
respectively.
They are defined as follows,
\begin{eqnarray}
 \mbox{\boldmath $u$}=(\rho \Gamma,\rho h \Gamma^2 v^1,\rho h \Gamma^2 v^2,
\rho h \Gamma^2-p-\rho \Gamma)^{T},\\
\mbox{\boldmath $f$}(\mbox{\boldmath $u$})=
(\rho \Gamma v^{1},\rho h\Gamma^2v^{1}v^{1}+p,
\rho h\Gamma^2v^{1}v^{2},
\rho h\Gamma^2 v^{1}-\rho \Gamma v^{1})^{T},\\
\mbox{\boldmath $g$}(\mbox{\boldmath $u$})=
(\rho \Gamma v^{2},\rho h\Gamma^2v^{1}v^{2},
\rho h\Gamma^2 v^{2}v^{2}+p,
\rho h\Gamma^2 v^{2}-\rho \Gamma v^{2})^{T},\\
\mbox{\boldmath $s_c$}
=\left(0,{p\over r},0,0\right)^T,\\
\mbox{\boldmath $s_s$}
=\left(0,{2p +\rho h \Gamma^2 v^2v^2 \over r},
-{\rho h \Gamma^2 v^1v^2-\cot \theta p\over r},
0\right)^T.
\end{eqnarray}
In this appendix, we use the unit that speed of light is unity.
We employ equation of state with constant adiabatic index
($p=(\gamma -1)\rho \epsilon$).
The discretized formula for the version of
the first order accuracy in time is as follows,
\begin{eqnarray}
\label{1storder}
\mbox{\boldmath $u$}^{n+1}
=\mbox{\boldmath $u$}^{n}-
\Delta tL(\mbox{\boldmath $u$}^{n}).
\end{eqnarray}
where subscripts indicate time steps, the time step is $\Delta t$,
and
a function of $L(\mbox{\boldmath $u$})$ is defined as,
for example in cylindrical coordinate case,
\begin{eqnarray}
L(\mbox{\boldmath $u$})\equiv\nonumber\\
{1\over \Delta r_i}
\{r_{i+1/2,j}\mbox{\boldmath $\tilde f$}(\mbox{\boldmath $u$})_{i+1/2,j}
+r_{i-1/2,j}\mbox{\boldmath $\tilde f$}(\mbox{\boldmath $u$})_{i-1/2,j}\}
+{1\over \Delta z_j}
\{\mbox{\boldmath $\tilde g$}(\mbox{\boldmath $u$})_{i,j+1/2}
-\mbox{\boldmath $\tilde g$}(\mbox{\boldmath $u$})_{i,j-1/2}\}
+\mbox{\boldmath $s_c$}(\mbox{\boldmath $u$})_{i,j},
\end{eqnarray}
where subscripts stand for space.
Time integration is carried out using TVD Runege-Kutta method \citep{Shu89}
to get higher order accuracy in time.
For example, we use following formulae in
the versions of the second and third order accuracy in time.
At first common process is done,
\begin{eqnarray}
\mbox{\boldmath $u$}^{(1)}=
\mbox{\boldmath $u$}^{n}+\Delta tL(\mbox{\boldmath $u$}^{n}).
\end{eqnarray}
$\mbox{\boldmath $u$}^{(1)}$ is assigned to be
$\mbox{\boldmath $u$}^{n+1}$ for the version of first order accuracy in
time
as shown in Eq.\ref{1storder}.
Following additional approaches are done
for the version of the higher order accuracy in time.

RK2 (2nd order accuracy)
\begin{eqnarray}
\mbox{\boldmath $u$}^{n+1}=
{1\over 2}
(\mbox{\boldmath $u$}^{(1)}+\Delta tL(\mbox{\boldmath $u$}^{(1)}))
\end{eqnarray}

RK3 (3rd order accuracy)
\begin{eqnarray}
\mbox{\boldmath $u$}^{(2)}=
{3\over 4}\mbox{\boldmath $u$}^{n}
+{1\over 4}
(\mbox{\boldmath $u$}^{(1)}+\Delta tL(\mbox{\boldmath $u$}^{(1)}))\\
\mbox{\boldmath $u$}^{n+1}=
{1\over 3}\mbox{\boldmath $u$}^{n}
+{2\over 3}
(\mbox{\boldmath $u$}^{(2)}+\Delta tL(\mbox{\boldmath $u$}^{(2)}))
\end{eqnarray}
The recovery from
conservative values $\mbox{\boldmath $u$}$
to primitive variables, such as, $\rho$, $p$, $\mbox{\boldmath $v$}$ is
done following \citet{Aloy99}.

Our code is based on the Godunov type scheme which is applied in
many relativistic hydrodynamic codes
and has an advantage to catch the shocks clearly compared with
other schemes.
The piecewise constant,
Monotone Upwind Scheme for Conservation Laws (MUSCL), and
PPM are adopted
for reconstruction to derive the cell surface states.
The MUSCL reconstruction used in the code was described
in \citet{Mizuta04}.
We followed the PPM reconstruction as shown by \citet{Marti96},
see also original paper of the PPM by \citet{Collela84}
for non-relativistic hydrodynamic code.
Those reconstruction can allow us to get
1st, 2nd, and 3rd order accuracy in space, respectively.

Numerical fluxes, such as,
$\mbox{\boldmath $\tilde f$}(\mbox{\boldmath$u$})$
and 
$\mbox{\boldmath $\tilde g$}(\mbox{\boldmath$u$})$
are
derived from the Marquina's flux formula \citep{Donat96,Donat98}
instead of using an exact Riemann solver
as \citet{Collela84} and \citet{Marti96} did.
The expressions of the eigenvalues and left and right eigenvectors
of Jacobian matrices, such as,
$\partial \mbox{\boldmath $f$}/\partial \mbox{\boldmath $u$}$ and
$\partial \mbox{\boldmath $g$}/\partial \mbox{\boldmath $u$}$
which are necessary to calculate numerical fluxes
are given in \citet{Donat98}.

Recent development of not only relativistic hydrodynamic code but also
relativistic magneto-hydrodynamic code allow
us to simulate high energetic phenomena,
such as relativistic jets from AGN, quasars, and micro quasars,
pulsar wind, and GRBs.
The approach to relativistic hydrodynamic simulations
by Godunov type scheme was employed in 90's.
\citet{Marti96} developed the PPM code
which uses an exact Riemann solver of relativistic hydrodynamics.
\citet{Eulderink95} derived the Roe average
for the general relativistic hydrodynamic equations and
extended the original Roe scheme, which is for non-relativistic hydrodynamics,
to general relativistic hydrodynamics.
Recently a number of codes have been developed
based on Godunov-type scheme which uses an approximate
Riemann solver to derive the numerical fluxes
for not only relativistic hydrodynamics
(\citet{Duncan94,Font94,Donat98,Delzanna02,Lucas04,Mignone05,Zhang05,Rahman05}),
but also relativistic magneto-hydrodynamics
(\citet{Komissarov99,Balsara01,Gammie03,Leismann05,Shibata05,Anton06}),
see also useful review by \citet{Marti03}.
The codes which do not use an exact or approximate Riemann solver
have also been developed by \citet{vanPutten91,Koide03,DeVilliers03}.

The results of the following all test calculations
are done by the version of the PPM reconstruction
and a third order accuracy version by TVD Runge-Kutta for time
integration.
The same version is used for our main results, i.e,
2D outflow propagation in the collapsar.
The detailed analysis of test results,
such as the error by different accuracy and resolution
will be presented in the paper in prep.

\subsection{1D shock tube problem with no transverse velocity}
\label{shtubetrans}
The shock tube problem is one of the standard test problems for
not only the compressive non-relativistic hydrodynamic codes
but also relativistic hydrodynamic codes,
since we can test the features of shocks,
rarefactions, and contact discontinues by this problem
and the analytic solutions are available via iterative calculations.
The problem is a kind of Riemann problem.
Two half finite constant states are assumed as an initial condition
($t=0$).
We have tested two standard sets of initial parameters
which were tested by the most of relativistic hydrodynamic codes.
These cases do not include any transverse velocity.
Those are shock tube problems A and B, and detailed parameters,
geometry and total grid points are :
\begin{itemize}
\item Shock tube A (plane) : 400 uniform grid points\\
Left \ \  state ($0<x<0.5$); $\rho_{L}=10, p_{L}=13.3, {v_{x}}_{L}=0,
      {v_{y}}_{L}=0, \gamma_{L}=5/3$\\
Right state ($0.5<x<1$); $\rho_{R}=1, p_{R}=1\times 10^{-6},
{v_{x}}_{R}=0, {v_{y}}_{L}=0,\gamma_{R}=5/3$\\

\item Shock tube B (plane) : 400 uniform grid points\\
Left \ \  state ($0<x<0.5$); $\rho_{L}=1, p_{L}=1000, {v_{x}}_{L}=0, 
{v_{y}}_{L}=0, \gamma_{L}=5/3$\\
Right state ($0.5<x<1$); $\rho_{R}=1, p_{R}=0.01,
{v_{x}}_{R}=0, 
{v_{y}}_{R}=0, \gamma_{R}=5/3$\\
\end{itemize}

\begin{figure}
\epsscale{.8}
\plotone{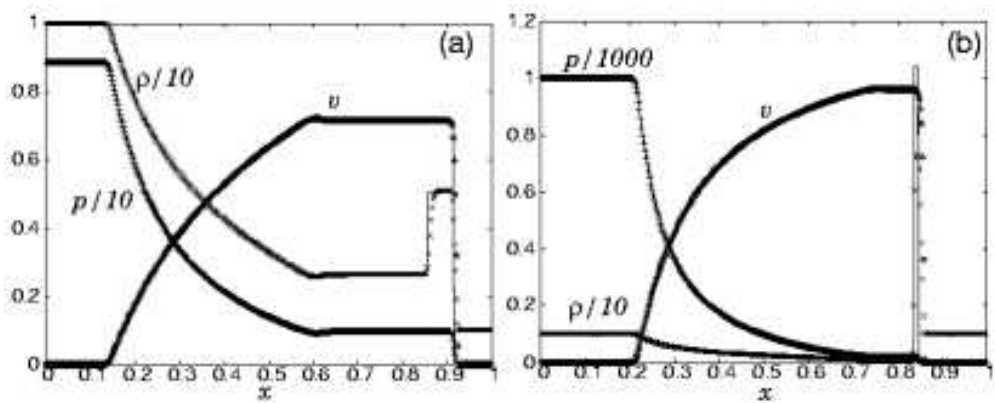}
\caption{Numerical (points) and analytic (solid lines)
solutions of 1D shock tube
problem without transverse velocity cases:
shock tube A (left) at $t=0.5$ and shock tube B (right) at $t=0.35$.
Density, pressure and velocity profiles are shown.
Uniform 400 zones are used for the calculation.}
\label{shocktube}
\end{figure}

Figure \ref{shocktube} shows numerical and analytic solutions
of these problems at $t=0.5$ (shock tube A) and
$t=0.35$ (shock tube B).
The profiles of the density, pressure, and 3-velocity are presented.
The numerical results of both problems are quite good agreement
with the analytic solution,
especially rarefactions.
In case A, shock front is captured in good accuracy
by several grid points.
The contact discontinuity is a little diffusive.
In case B, because of large pressure jump of initial condition,
the jump of the density is
about 10 times at the shock
and very narrow shocked region appears.
It is hard to resolve such narrow region by tested resolution.
But our results are within the similar level of
the results presented by other groups.

\subsection{1D shock tube problem with transverse velocity}
\label{SHtranseverse}
\begin{landscape}
\begin{figure}
\epsscale{.7}
\rotatebox{270}
{\plotone{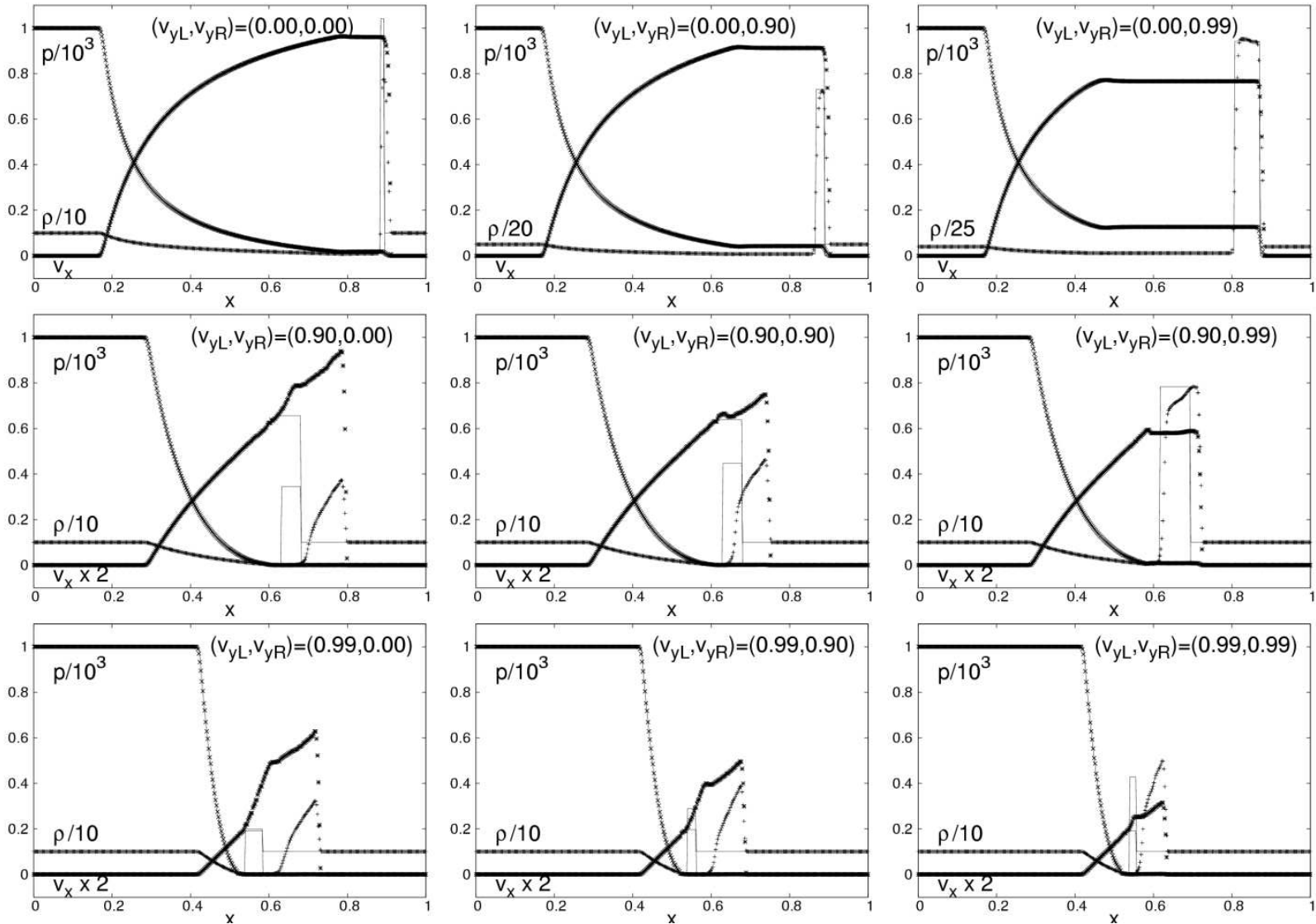}}
\caption{Numerical (points) and analytic (solid lines)
solutions of shock tube
problem with transverse velocity case with uniform 400 zones
(shock tube C) at $t=0.4$
are presented.
From left to right, ${v_{y}}_R=0,0.9,0.99$
and from top to bottom ${v_{y}}_L=0,0.9,0.99$.
Density, pressure, x-component of velocity are shown.}
\label{shocktubeMI}
\end{figure}
\end{landscape}

If there is non-zero transverse velocity for 1D shock tube problem,
the results changes because of the dependence
of the Lorentz factor on the absolute value of the velocity.
as discussed by \citet{Pons00,Rezzolla01}.
Recently it has been reported
that it is numerically hard to resolve such problems with
as same level of grid points as the cases without transverse velocities
\citet{Mignone05,Zhang05}.
We followed the conditions by \citet{Mignone05}
in which the effect of some different initial transverse velocities
cases was presented.
The initial conditions are:
\begin{itemize}
\item Shock tube C (plane) uniform 400 grid points\\
Left \ \  state ($0<x<0.5$); $\rho_{L}=1.0, p_{L}=1000.0, {v_{x}}_L=0,
      \gamma_{L}=5/3$\\
Right state ($0.5<x<1$); $\rho_{R}=1.0, p_{R}=1\times 10^{-2}, {v_{x}}_R=0,
\gamma_{R}=5/3$
\end{itemize}
We have done parametric study by changing
${v_y}_L$ from 0 to 0.99 and 
${v_y}_R$ from 0 to 0.99 as  \citet{Mignone05} did.
Figure \ref{shocktubeMI} shows the results of these tests.
The profiles of the
density, pressure and $x$-component of the 3-velocity
at $t=4.0$ are presented.
The top-left panel (case (${v_y}_L, {v_y}_R$)=(0,0))
corresponds to the case Shock tube B presented above,
but the result at different time is shown.
The rarefactions are resolved in any cases as shown
the cases without transverse velocity (shock tube A and B).
As ${v_y}_L$ increases, both positional and absolute value
errors at around the discontinuities, such as,
shocks and contact discontinuities increase.

\begin{figure}
\begin{center}
\epsscale{1.2}
\rotatebox{270}
{\plotone{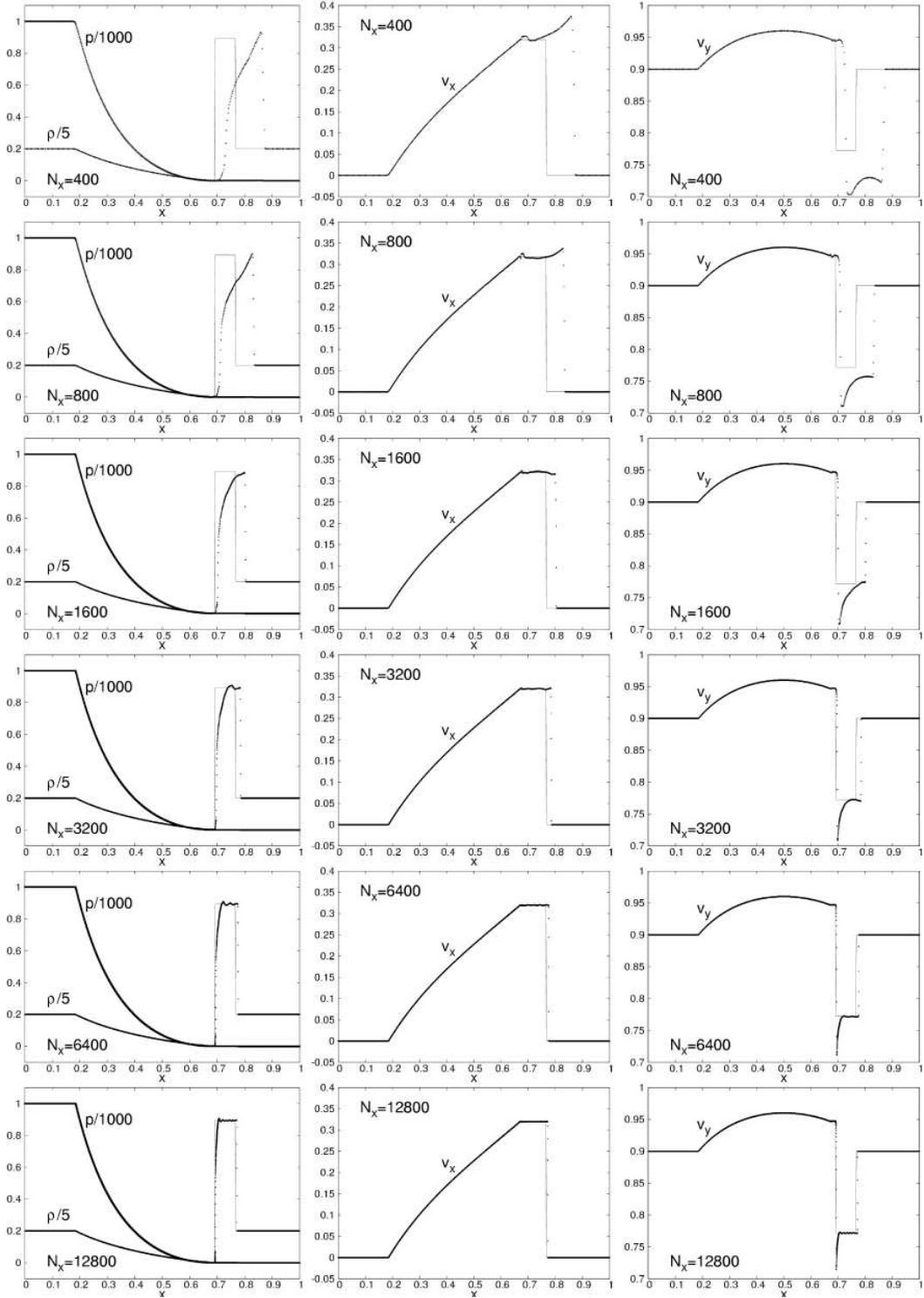}}
\caption{Numerical (points) and analytic (solids) solutions of shock tube
problem with transverse velocity case
(shock tube C, (${v_y}_L, {v_y}_R$)=(0.9,0.9)):
Different resolution of the calculations are presented,
from top to bottom, the number of grid points are
400, 800, 1600, 3200, 6400, 12800, respectively.
Density, pressure and x- and y-velocity components are shown.}
\label{shocktubetrans}
\end{center}
\end{figure}

We also done resolution study on the set of
(${v_y}_L, {v_y}_R$)=(0.9,0.9)
from uniform 400 zones to 12800 zones.
This has been done by \citet{Zhang05}
but they used their Adaptive Mesh Refinement (AMR) version for this.
Figure \ref{shocktubetrans} shows numerical and analytic solutions
of this problem.
The rarefaction is resolved in good accuracy
by not only higher resolution calculation but also lower one.
On the contrary, both the shock and the contact discontinuities
are not resolved in both position and absolute value
by lower resolution calculations.
Higher resolution calculations can allow us to resolve
shock front within a few percents error in position.
On the contrary,
there still remains about several percents error in $v_y$ at
the contact discontinuity.
One of solutions to avoid such errors and to get good accuracy
is to employ the AMR method
which uses locally higher resolution at the area
where discontinuities, such as, shocks and contact discontinuities
exist to save the CPU time and memory requirement.
The AMR was employed by \citet{Zhang05} is one of the ways
to reduce the error seen in this problem.

\subsection{1D Reflection Shock Problem}
Reflection shock problem is suitable to test strong shocks.
At $t=0$,
an uniform, high Mach number, and cold flow reflects at the boundary
($x=0$ or $r=0$)
which is a wall for the Cartesian coordinate case,
a cylindrical axis for the cylindrical coordinate case,
and center of the spherical coordinate for the spherical coordinate case.
A strong shock appears and proceeds to the cold gas.
The fluid is heated by this shock and becomes at rest.
The shock front satisfies
Rankine-Hugoniot relation for relativistic hydrodynamics \citep{Taub48}.
An analytic solution is available \citep{Johnson71},
if the gas of the initial flow is cold ($p_0\sim 0$).

The analytic solution at $t=T$ is,
\begin{eqnarray}
\rho=\left({\gamma+1\over \gamma-1}
+{\gamma\over \gamma-1}(\Gamma_0-1)\right)\rho_0,
\epsilon =\Gamma_0-1, v=0,\ \mbox{for}\  x< V_s T \mbox{\ or\ } r< V_s T \\
\rho=\rho_0\left(1+{|v_0|t\over x}\right)^{\alpha},\epsilon \sim 0,v=v_0,
\ \mbox{for} \ x>V_{s} T \mbox{\ or\ } r>V_{s} T,
\end{eqnarray}
where $V_s$ is shock velocity,
\begin{eqnarray}
V_{s}=\left({\Gamma_0-1\over \Gamma_0+1}\right)^{1/2}(\gamma-1).
\end{eqnarray}
The rest mass density, specific internal energy,
velocity, and Lorentz factor of the inflow are
$\rho_0$, $\epsilon_0$, $v_0$, and $\Gamma_0$, respectively.
The geometry is plane ($\alpha =0$), cylindrical ($\alpha =1$),
and spherical ($\alpha =2$), respectively.
The expression of rest mass density jump at the shock front is
divided in two parts, namely
a maximum density compression ratio in non relativistic hydrodynamics
($(\gamma+1)/(\gamma-1)$) and including a Lorentz factor.

\begin{itemize}
\item REP\\
$\rho_0=1.0,\epsilon_0=10^{-4},v_0=-0.999(\Gamma_0=22), \gamma =4/3$, Plane

\item REC\\
$\rho_0=1.0,\epsilon_0=10^{-4},v_0=-0.999(\Gamma_0=22), \gamma =4/3$, Cylindrical

\item RES\\
$\rho_0=1.0,\epsilon_0=10^{-4},v_0=-0.999(\Gamma_0=22), \gamma =4/3$, Spherical
\end{itemize}

\begin{figure}
\begin{center}
\epsscale{1.}
\rotatebox{270}
{\plotone{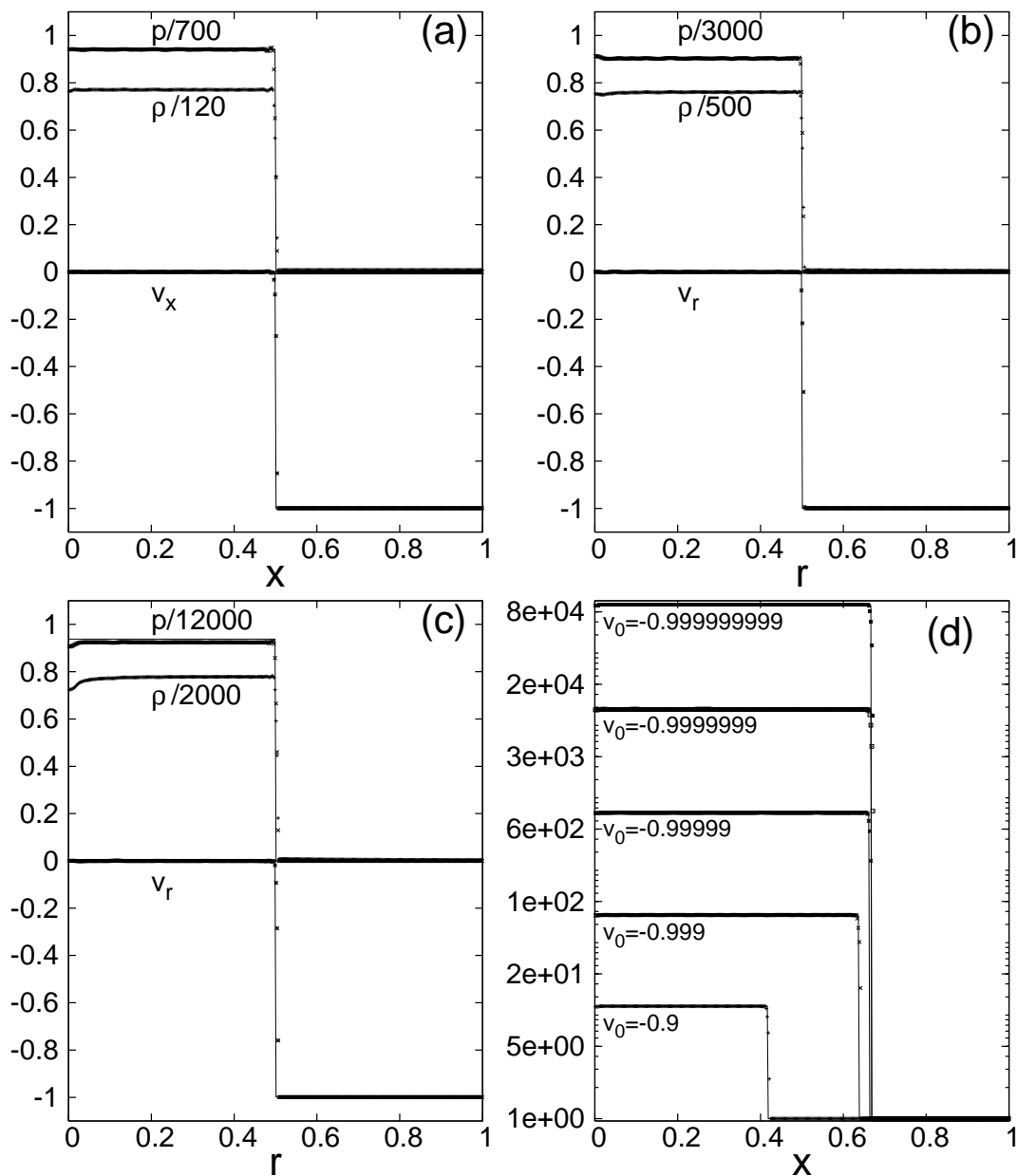}}
\caption{Numerical (points) and analytic (solid lines)
solutions of reflection shock problem with 400 uniform zones.
The calculations are done in
(a) plane geometry, (b) cylindrical geometry,
(c) spherical geometry ($v_0=-0.99$).
The results at $t=1.57$ are presented.
The results of rest mass density for different initial velocity from
$v_0=-0.9$ to $v_0=-0.999999999$ at $t=2.0$ are shown in (d).}
\label{reflection}
\end{center}
\end{figure}

Figure \ref{reflection} (a)-(c) show numerical and analytic solutions
of this problem at $t=1.57$.
The calculations are done uniform 400 zones in the Cartesian (a),
 cylindrical (b),
and  spherical coordinate (c), respectively.
Since we have taken care of the treatment of the boundary
for the cylindrical and spherical coordinate cases,
the error around the boundary (singular) is reduced
compared with the results shown in \citet{Mizuta04}.
The error at and round the wall boundary
is studied by \citet{Noh87} in detail.
He discussed not only case in plane geometry but also
cylindrical and spherical geometry. 
Our numerical error at the wall is 0.8 \% (Cartesian),
1.4\% (cyrindrical), and 8.3\% (spehrical).

Figure \ref{reflection} (d) shows numerical and analytic solutions
of reflection shock problem by Cartesian coordinate, but
different initial velocity,
from $-0.9$ up to $-0.999999999$,
corresponding Lorentz factor is from 2.27 to 22361.
Uniform 400 zones are used.
The other conditions, such as, density and pressure is as same as
the problem (REP).
In all cases, the strong shock front is captured with
a few grid points.
All results are at $t=2$.
The jump condition at the shock and location of the shock
represents analytic solution.

\subsection{2D shock tube problem}
\begin{figure}
\epsscale{.8}
\plotone{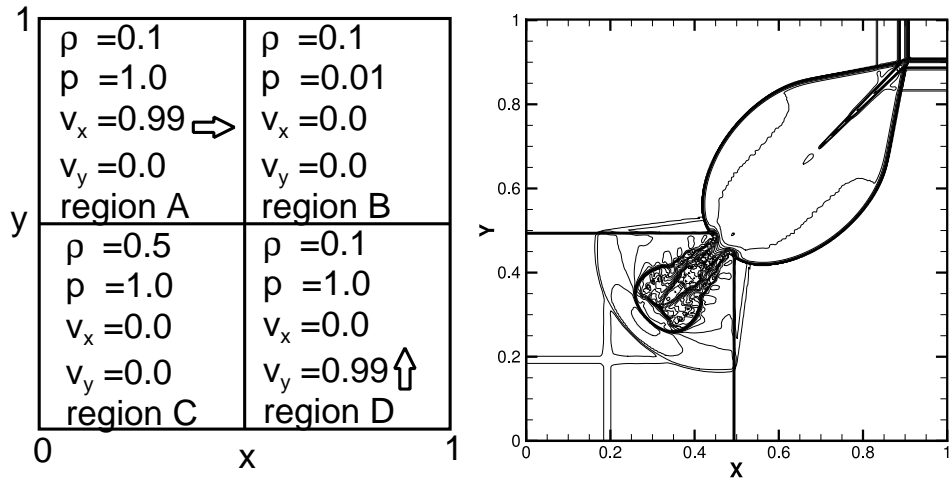}
\caption{Schematic figure of the initial condition of 2D shock tube
problem (left) and numerical results (right).
uniform $400\times 400$ zones are used.
30 levels of iso-surface of the logarithm
of rest mass density at $t=0.4$ is shown.}
\label{2dshtube}
\end{figure}

Two dimensional shock tube problem is done to confirm
the shock dynamics in the multidimensional case.
This problem includes the interactions of shocks,
rarefactions, contact discontinuities.
Initially a square computational domain is
prepared in x-y plane and divided into four quarter boxes
(see left panel in Fig.\ref{2dshtube}).
Initial conditions in each box are :
\begin{eqnarray}
(\rho,v_x,v_y,p)=(0.10,0.00,0.00,0.01)
: 0.5\le x\le 1, \ 0.5\le y\le 1 \ (\mbox{region A}),\nonumber \\
(\rho,v_x,v_y,p)=(0.10,0.99,0.00,1.00)
:  0\le x\le 0.5, \ 0.5\le y\le 1 \ (\mbox{region B}) ,\nonumber \\
(\rho,v_x,v_y,p)=(0.50,0.00,0.00,1.00)
: 0\le x\le 0.5, \ 0\le y\le 0.5 \ (\mbox{region C}),\nonumber \\
(\rho,v_x,v_y,p)=(0.10,0.00,0.99,1.00)
: 0.5\le x\le 1, \ 0\le y\le 0.5 \ (\mbox{region D}).\nonumber
\end{eqnarray}
Adiabatic index is assumed to be $\gamma =5/3$.
This set of initial condition is as same as used by
\citet{Delzanna02,Zhang05}.
This is also similar condition done by \citet{Lucas04}.
At first two shocks appear from the high pressure
regions A and D and they proceed into the region B.
These shocks collide each other in the region B.
As a result, a high pressure gas break into the region C.
We use $400\times 400$ uniform grid points
in a square computational box.
The boundary conditions at all boundary are open ones.

Figure \ref{2dshtube} shows
30 levels of iso-surface of the logarithm of rest mass density
at $t=0.4$.
Two curve of shocks in the region B are well resolved
as shown by other groups.
It should be noted that
sound waves appear at the contact discontinuities
between the region A and C, and the region C and D
can be seen in the density contour plot in the region C.
Such wave can also be seen in the results
\citet{Delzanna02,Lucas04,Zhang05}.

\subsection{2D Double Mach Reflection Problem}
2D double Mach reflection problem was introduced by \citet{Woodward84} for
non-relativistic hydrodynamics
and recently has been applied to relativistic
hydrodynamics by \citet{Zhang05}.
An initial uniform shock collides with a reflective wall,
forming another shock.
These two shocks interact each other.
A kind of jet appears along the wall.
The solution is self similar.

We have tested this problem,
using as same parameters as \citet{Zhang05} did.
The density and pressure of the unchecked gas is
1.4 and 0.0025, respectively.
The classical shock Mach number, $M_{c}\equiv v_S/c_S$,
where $v_S$ is shock velocity and $c_{S}$
is sound velocity of the unchecked gas, respectively, is 10.
The adiabatic index is $\gamma =1.4$.
The shock velocity in this case is $\sim 0.4984$ accordingly.
The state of initial shocked gas
can be derived by Rankine-Hugoniot relation
of relativistic hydrodynamics.
The density, pressure, and velocity of shocked gas
are $\sim 8.564$, $\sim 0.3808$, and $\sim 0.4247$, respectively.
The computational domain is $x-y$ plane and $4\times 1$ rectangular
which includes uniform $512\times 128$ zones.
The shock front makes an angle of 60 degrees with $x$ axis
and the rim of the shock is at $x=1/6,\ y=0$ initially.
The boundary conditions at $x=0$, $0<x<1/6$ at $y=0$,
and $0<x<V_s\sin 60^\circ\ t$ at $y=1$ are
inflow of the initial shocked gas.
The reflective boundary condition is employed at $1/6<x<4, y=0$.
The boundary condition at other boundaries are open condition.

\begin{figure}
\epsscale{1.0}
\plotone{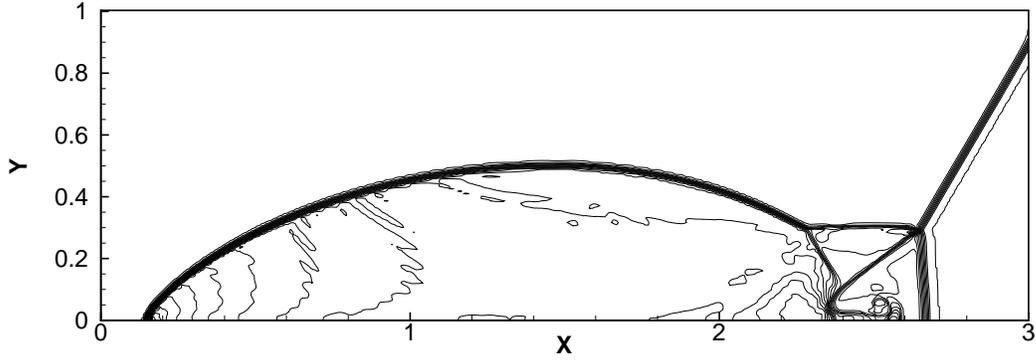}
\caption{30 levels of iso-surface of the rest mass density of
double Mach shock at $t=4.0$.
$512\times 128$ uniform zones and adiabatic index $\gamma=1.4$
for equation of state are used.}
\label{double}
\end{figure}

Figure \ref{double} shows
30 levels of iso-contours of the rest mass density at $t=4.0$. 
The global feature of the shocks is represented as shown by
\citet{Zhang05}.
Although the resolution of presented calculation is not so high,
the Kelvin-Helmholtz instability
which violates the self similarity of the solution
can be seen at around $x=2.5,\ y=0$.

\subsection{2D Emery Step Problem}
The problem of a wind tunnel containing a step 
was proposed by \citet{Emery68} has been tested
(2D Emery problem).
He tried to compare the results by different schemes.
A step boundary is introduced
in the rectangular computational box ($x\times y=3\times 1$).
This step is located at 0.6 from left boundary of the
computational box and the hight is 0.2
and continues to the right boundary.
The left side boundary is an inflow as the same quantities.
Initially computational box is filled with
a uniform supersonic flow
($v_x=0.999$, $v_y=0$, $\rho=1.4$, and Newtonian Mach number is 3).
The outflow (zero gradient) boundary condition is employed
at right side of the boundary ($x=3$).
The reflective boundary conditions are employed
at the step boundaries, $y=1$, and $0<x<0.6$ at $y=0$.
$120\times 40$ uniform grid points are spaced.
The adiabatic index for equation of state is $\gamma=1.4$.
The corner of the step becomes singular.
The corrections around the corner of the step
have been included, see Appendix in \citet{Donat98}.

Figure \ref{emery} shows 30 levels of iso-surface of the logarithm of
rest mass density.
The global features, such as shocks and rarefactions,
are reproduced as shown in \citet{Zhang05}.

\subsection{Spherical Blast Wave}
At last, we show the results of spherical blast wave,
using cylindrical coordinate.
This problem is done to see how spherical symmetry
is kept using cylindrical coordinate.
The computational box is $0<z<1, 0<r<1$
with equally spaced grid points ($320\times 320$).
Initially, a high pressure gas ($p=1000$) is put in
the region, $\sqrt{z^2+r^2}\le 0.4$.
The pressure in the outside is  $p=1$.
At $t=0$ all gas is at rest and the density is uniform $\rho=1$.
The adiabatic index for equation of state is $\gamma=5/3$.
The high pressure gas expands and a shock proceeds into cold gas.
The boundary conditions are reflective one at both axises
and zero gradient outflow one at the other boundaries.
The same problem has been done using
spherical coordinate with 3200 uniform zones
for comparison.

\begin{figure}
\begin{center}
\epsscale{.9}
\rotatebox{270}
{\plotone{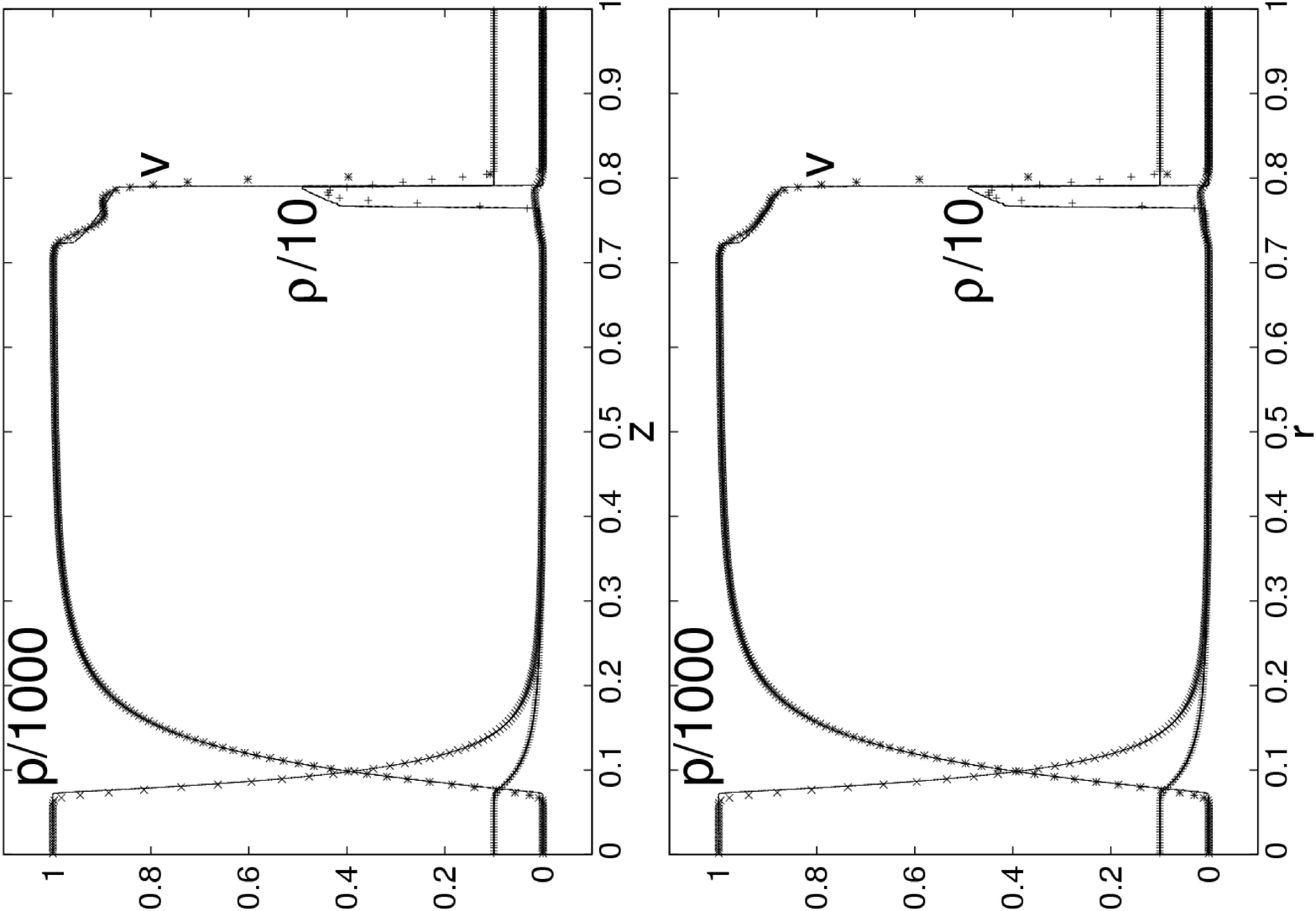}}
\caption{Spherical blast wave problem by
cylindrical coordinate with uniform $320\times 320$ ($r\times z$) zones.
Rest mass density, pressure, velocity along the axis at $t=0.4$
are presented.
Top: along $z$ axis. Bottom: along $r$ axis.
Solid lines are results of the same problem but
done by spherical coordinate with uniform 3200 zones.}
\label{blast}
\end{center}
\end{figure}

Figure \ref{blast} shows one dimensional rest mass density profile
of along both $z$ (top) and $r$ (bottom) axises at $t=0.4$.
The solid lines are the results of the same problem
by spherical coordinate with 3200 uniform zones in radius.
The both results along $z$ and $r$ axises are in good agreements with
spherical one,
although the velocity profile along $z$ axis has
a bump around the head of the wave $x \sim 0.75$.

\end{appendix}

\clearpage

\begin{table}[t]
\begin{center}
\caption{Calculated models.
The basic model parameters are the
Lorentz factor ($\Gamma_0$) and the specific internal 
energy ($\epsilon_0$).
Also shown are the ratio ($f_0$) 
of the kinetic energy to the total energy
(where rest mass energy is excluded) and the
estimated maximum Lorentz factor ($\Gamma_{\rm max}$).
The achieved maximum Lorentz factor ($\Gamma_{\rm max,sim}$)
and half opening angle ($\theta_{\rm sim}$)
from the calculations are also listed.}

\begin{tabular}{ccc|cc|ccc}
model & $\epsilon_0/c^2$ & {$\Gamma_0$ ($v_0/c$)} & $f_0$ &
 ${f_0}^{-1}$ & $\Gamma_{\rm max}$ & $\Gamma_{\rm max,sim}$ & $\theta_{\rm sim}$ \\
\hline \hline
A01  & 0.1 & & 0.86 & 1.2 & 5.5 & 5.6 & 1.2 \\
A10  & 1.0 & 5 (0.98) & 0.38 & 2.7 & 10 & 11 & 1.2 \\
A50  & 5.0 & & 0.11 & 9.3 & 30 & 31 & 1.7 \\ 
A300  & 30 & & 0.020 & 51 & 155 & 104 & 1.0 \\ \hline
B01  & 0.1 & & 0.85 & 1.2 & 4.4 & 4.5 & 1.6  \\
B10  & 1.0 & 4 (0.97) & 0.36 & 2.8 & 8.0 & 9.0 & 1.7 \\
B50  & 5.0 & & 0.10 & 9.8 & 24 & 26 & 1.3 \\ \hline
C01  & 0.1 & & 0.84 & 1.2 & 3.3 & 3.4 & 1.5 \\
C10  & 1.0 & 3 (0.94) & 0.34 & 2.9 & 6.0 & 6.7 & 1.8 \\
C50  & 5.0 & & 0.093 & 11 & 18 & 20 & 1.9 \\ \hline
D01  & 0.1 & & 0.80 & 1.3 & 2.2 & 2.3 & 2.9\\
D10  & 1.0 & 2 (0.87) & 0.29 & 3.5 & 4.0 & 4.5 & 15 \\
D50  & 5.0 & & 0.074 & 14 & 12 & 14 & 12 \\ \hline
E01  & 0.1 & & 0.71 & 1.4 & 1.5 & 1.6 & 12 \\
E10  & 1.0 & 1.4 (0.7) & 0.20 & 5.1 & 2.8 & 3.1 & 20 \\
E50  & 5.0 & & 0.0470 & 21 & 8.4 & 9.4 & 19 \\ \hline
F01  & 0.1 & & 0.64 & 1.6 & 1.4 & 1.4 & 19 \\
F10  & 1.0 & 1.25 (0.6) & 0.15 & 6.6 & 2.5 & 2.8 & 28 \\
F50  & 5.0 & & 0.034 & 29 & 7.5 & 8.3 & 26 \\ \hline
G01  & 0.1 & & 0.55 & 1.8 & 1.3 & 1.4 & 22 \\
G10  & 1.0 & 1.15 (0.5) & 0.11 & 9.1 & 2.3 & 2.6 & 30 \\
G50  & 5.0 & & 0.024 & 41 & 6.9 & 7.7 & 26 \\ \hline
H01  & 0.1 & 1.1 (0.4) & 0.44 & 2.3 & 1.2 & 1.3 & 27 \\
H10  & 1.0 & & 0.073 & 14 & 2.2 & 2.5 & 29 \\ \hline
I01  & 0.1 & 1.05 (0.3) & 0.31 & 3.2 & 1.2 & 1.2 & 32
\end{tabular}
\label{conditions}
\end{center}
\end{table}

\clearpage

\clearpage

\clearpage

\clearpage

\clearpage

\clearpage

\clearpage

\clearpage

\clearpage

\clearpage
\clearpage

\clearpage

\clearpage

\clearpage

\begin{figure}
\epsscale{.9}
{\plotone{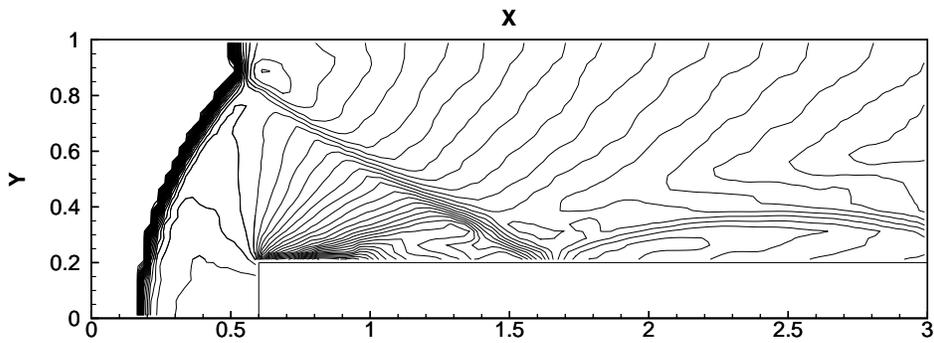}}
\caption{Emery step at $t=4.0$ with $120\times40$ uniform zones.
30 levels of iso-surface of the logarithm of rest mass density 
is shown.}
\label{emery}
\end{figure}

\end{document}